\documentclass{JHEP3}

\usepackage{latexsym}
\usepackage{graphics}

\def\comment #1{}
\def\cf {{\it cf. }}

\def\refer #1{{(\ref{#1})}}
\def\fullref #1{\ref{#1} (p.\pageref{#1})}

\def\bra #1{\left\langle{#1}\right|}
\def\ket #1{\left|{#1}\right\rangle}

\def\of #1{\!\left({#1}\right)}

\def\N {\mathrm{I\!N}}
\def\R {\mathrm{I\!R}}
\def\C {\mathrm{\,I\!\!\!C}}

\def\Im {\mathrm{Im}}

\def\grad {\nabla}
\def\gradOp {\hat\grad}

\def\set #1{\left\lbrace{#1}\right\rbrace}

\def\brackets #1{\left[{#1}\right]}
\def\braces #1{\left\lbrace{#1}\right\rbrace}

\def\order #1{\mathcal{O}\of{#1}}

\def\commutator #1#2{\brackets{{#1},{#2}}}
\def\antiCommutator #1#2{\braces{{#1},{#2}}}

\def\defas {:=}
\def\shallbe {\stackrel{!}{=}}
\def\equalby #1{\stackrel{\refer{#1}}{=}}

\def\manifold {\mathcal{M}}

\def\extd {\mathbf{{d}}}

\def\eigenspace #1#2 {\mathrm{eig}\of{#1,#2}}

\newlength{\skiplength}

\title{Nonabelian 2-forms and loop space connections from SCFT deformations}
\author{Urs Schreiber \\ Universit{\"a}t Duisburg-Essen \\ Essen, 45117, Germany\\
   E-mail: \email{Urs.Schreiber@uni-essen.de}}

\abstract{It is shown how the deformation of the superconformal generators on the
string's worldsheet
by a nonabelian super-Wilson line gives rise to a covariant exterior derivative on loop
space coming from a nonabelian 2-form on target space. The expression obtained this way is new
in the context of strings (but has been considered before in the context of integrable systems),
and its consistency is verified by checking that its global gauge transformations on loop space
imply the familiar gauge transformations on target space. We
derive the second order gauge transformation from infinitesimal local gauge transformations on loop space and find that 
a consistent picture is obtained only when the sum of the 2-form and the 1-form field strengths vanish.
The same condition has recently been derived from 2-group gauge theory reasoning. 
We observe that this condition implies that the connection on loop space is \emph{flat},
which is a crucial sufficient condition for the nonabelian surface holonomy induced by it
to be well defined.
Finally we compute the background equations of motion of the nonabelian 2-form by
canceling divergences in the deformed boundary state.}

\begin{document}

\newpage

\section{Introduction}

The target space theories which give rise to non-abelian 2-forms are not at all well understood
\cite{Ganor:1996}.
One expects \cite{BekaertHenneauxSevrin:2000,Hofman:2002} 
that they involve stacks of 5-branes on which open membranes may end
\cite{Lechner:2004,BraxMourad:1997,Strominger:1995}. This has recently been
made more precise \cite{AschieriJurco:2004}
using anomaly cancellation on M5-branes and the language of nonabelian gerbes
developed in \cite{AschieriCantiniJurco:2004}. 
The boundary of these membranes appear as strings, 
\cite{Witten:1997,Townsend:1996}, 
(self-dual strings \cite{CallanMaldacena:1997,Gibbons:1997,HoweLambertWest:1997},
``little strings'' \cite{Aharony:1999},
fundamental strings or D-strings 
\cite{LosevMooreSamsonShtashvili:1997}) 
in the world-volume theory of the 5-branes \cite{Strominger:1995}, generalizing \cite{Ganor:1996}
the way how open string endpoints appear as ``quarks'' in the world-volume theory of D-branes.
Just like a nonabelian 1-form couples to these ``quarks'', i.e. to the boundary of an open string,
a (possibly non-abelian) 2-form should couple
\cite{Kalkkinen:1999}
 to the boundary of an open membrane 
\cite{Townsend:1996,BraxMourad:1997b,KawamotoSasakura:2000,BergshoeffBermanVanDerSchaarSundell:2001}, 
i.e. a to string on the (stack of) 5 branes. 
One proposal for how such a non-abelian $B$ field might be induced by a stack of branes
has been made in \cite{Kalkkinen:1999}. A more formal derivation of the non-abelian
2-forms arising on stacks of M5 branes is given in \cite{AschieriJurco:2004}.
General investigations into the possible nature of such non-abelian 2-forms
have been done for instance in \cite{Lahiri:2001,Lahiri:2003}.

(From the point of view of the effective 6-dimensional
supersymmetric worldvolume theory of the 5-branes these 2-form field(s) come either
from a tensor multiplet or from a gravitational multiplet of the 
worldvolume supersymmetry representation \cite{SeibergWitten:1996}.)

This analogy strongly suggests that there is a \emph{single} Chan-Paton-like factor associated to
each string living on the stack of 5 branes, indicating which of the $N$ branes in the stack it 
is associated with. This Chan-Paton factor should be the degree of freedom that the non-abelian
$B$-field acts on.

Hence the higher-dimensional generalization of ordinary gauge theory should, in terms of strings,
involve the steps upwards the dimensional ladder indicated in table \ref{dimensional ladder}.

\TABULAR[h]{|lcl|}{
  \hline
  (1-)gauge theory && 2-gauge theory\\
  \hline\hline
  string ending on D-brane &$\to$& membrane ending on NS brane\\ \hline 
  ``quark'' on D-brane &$\to$& string on NS brane\\ \hline
  nonabelian 1-form gauge field &$\to$&  nonabelian 2-form gauge field $B$\\ \hline
  \parbox{6cm}{
  coupling to the boundary of a 1-brane (string)} 
   &$\to$& 
  \parbox{6cm}{coupling to the boundary of a 2-brane (membrane)}\\ \hline
  \parbox{6cm}{Chan-Paton factor indicating which D-brane in the stack the ``quark'' sits on}
  &$\to$&
  \parbox{6cm}{
  Chan-Paton-like factor indicating with NS brane in the stack the membrane boundary string sits on.}
  \\ \hline
}{\label{dimensional ladder} Expected relation between 1-form and 2-form gauge theory in stringy terms}

These considerations receive substantiation by the fact that 
indeed the contexts 
in which nonablian 2-forms have been argued to arise naturally are the
worldsheet theories on these NS 5-branes
\cite{Hofman:2002,BekaertHenneauxSevrin:2000,Witten:1997,LosevMooreSamsonShtashvili:1997,Ganor:1996,Witten:1995b}.

The study of little strings, tensionless strings and  $N=(2,0)$ QFTs in six dimensions is
involved and no good understanding of any non-abelian 2-form from this target space
perspective has emerged so far.
However, a compelling connection is the relation of these 6-dimensional theories, 
upon compactification, to 
Yang-Mills theory in 4-dimensions, where the 1-form gauge field of the Yang-Mills theory
arises as one component of the 2-form in the 6-dimensional theory \cite{Witten:1995b}.

For this to work the dimension $d=5+1$ of the world-volume theory of the 5-branes plays a crucial role,
because here the 2-form $B$ can have and does have \emph{self-dual} field strength $H = \star H$
\cite{Townsend:1996,Witten:1995b}
(related to the existence of the self-dual strings in 6 dimensions 
first discussed in \cite{DuffLu:1993}). 

But this means that there cannot be any ordinary non-topological action of the form $dH\wedge \star dH$
for the $B$-field, and that
furthermore the dynamical content of the $B$ field would essentially be that of a 1-form $\alpha$
\cite{Witten:1995b}: Namely when the 1+5 dimensional field theory is compactified on a circle
and $B$ is rewritten as 
\begin{eqnarray}
  B = B_{ij}\,dx^i \wedge dx^j + \alpha_i dx^i \wedge dx^6 \hspace{1cm}\mbox{for $i,j \in \set{1,2,3,4,5}$}
  \,.
  \nonumber
\end{eqnarray}
with $\partial_6 B = 0$, then $dB = \star dB$ implies that in five dimensions $B$ is just dual
to $\alpha$
\begin{eqnarray}
  d^{(5)}B = \star^{(5)} d\alpha
  \,.
\end{eqnarray}
In particular, since the compactified theory should give possibly non-abelian Yang-Mills with $\alpha$
the gauge field \cite{Witten:1995b} it is natural to expect 
\cite{Hofman:2002} that in the uncompactified theory there must be a non-abelian $B$ field.
Since there is no Lagrangian description of the brane's worldvolume theory 
\cite{Witten:1996,BekaertHenneauxSevrin:2000} 
it is hard to make this explicit. This is one reason why it seems helpful to consider the
worldsheet theory of strings propagating in the 6-dimensional brane volume, as we will do here.
The non-abelian
Yang-Mills theory in the context of NS 5-branes considerd in \cite{Witten:1997} uses $n$ D4-branes
suspended between two NS5-branes. The former can however be regarded as a single M5-brane wrapped $n$
times around the $S^1$ (\cf p. 34 of \cite{Witten:1997}).

In \cite{BermanHarvey:2004} it is argued that, while the worldvolume theory 
on a stack of 5-branes with non-abelian 2-form fields is not known, it cannot
be a local field theory. This harmonizes with the attempts in \cite{Ganor:1996}
to define it in terms of ``nonabelian surface equations'' which are supposed to generalize the
well-known Wilson loop equations of ordinary Yang-Mills theory to Wilson \emph{surfaces}.
These Wilson surfaces become ordinary Wilson loops in loop space, and these play a pivotal role
in the constructions presented here.

In the following we shall make no attempt to say anything \emph{directly} about the physics of membranes
attached to 5 branes. Instead the strategy here is to look at the worldsheet theory of strings 
and to see from the 
worldsheet perspective if anything can be said about superconformal field theories
that involve a non-abelian 2-form. Even though we will not
try to exhibit a direct correspondence between certain such SCFTs and 
the target space physics of membrane boundary strings on NS branes, we will be able
to recognize 
in the formal structure of these SCFTs
some of the above mentioned expected properties of such theories. In the process we
will clarify or shed new light on previous approaches to non-abelian surface holonomy
\cite{Hofman:2002,AlvarezFerreiraSanchezGuillen:1998}, in particular by deriving the
form of the loop space connection from SCFT deformations (boundary state deformations) and
deriving its crucial flatness condition from consistency conditions on its gauge invariance.
This flatness condition turns out to be already known \cite{GirelliPfeiffer:2004}
in the context of 2-group theory
\cite{Baez:2002}, and together with the form of the loop space connection
used here it is seen to solve the 
famous old problem noted \cite{Teitelboim:1986}, related to the construction of
a sensible notion of surface ordering for non-abelian 2-forms.
By merging the insights into non-abelian surface holonomy using the loop space results
found here and in \cite{AlvarezFerreiraSanchezGuillen:1998} with those
of 2-group theory, one obtains a coherent picture clarifying aspects of both
methods. This is reported in full detail in \cite{Schreiber:2004g}.\\

The aim of this paper is to show that 
it is possible to learn about the physics of strings in non-abelian 2-form backgrounds
by suitably generalizing 
known deformation techniques of 2d superconformal field theories from the abelian to the
nonabelian case, this way learning about the nonabelian target space background and
its effective nonabelian 2-form field theory from the study of appropriate worldsheet theories.

In particular, the technique of SCFT deformations using ``Morse theory methods''
in loop space formalism \cite{Schreiber:2004f} 
that was studied in \cite{Schreiber:2004}, together with insights into boundary state
deformations obtained in \cite{Hashimoto:2000,Hashimoto:1999},
is used to construct superconformal worldsheet generators that incorporate a 
connection on loop space induced by a nonabelian 2-form on target space. 

Given a nonabelian 1-form $A$ and a 2-form $B$ on target space (which we think of as taking
values in some matrix algebra) we argue from general SCFT deformation theory that the connection 
on loop space (the space of maps from the circle into target space) induced by this background
can be read off from a similarity transformation of the worldsheet supercharges $G$ and $\bar G$
by the operator
\begin{eqnarray}
  \label{deformation operator in introduction}
    \exp\of{\bf W}^{(A)(B)_\mathrm{nonab}}
  &=&
  \mathrm{P}
  \exp\of{\int\limits_0^{2\pi}
    d\sigma\;
    \left(
      i A_\mu X^{\prime\mu}
      +
      \frac{1}{2}
      \left(\frac{1}{T}F_A + B\right)_{\mu\nu} {\cal E}^{\dagger \mu}{\cal E}^{\dagger \nu}
    \right)
  }
  \,.
\end{eqnarray}
Here $X\of{\sigma}$ is the map from the loop into target space, $X^{\prime}(\sigma) =\frac{d}{d\sigma}X\of{\sigma}$
is the tangent vector,
 $T=1/2\pi \alpha^\prime$ is the string tension, $F_A$ the field strength of $A$
and ${\cal E}^\dagger$ are operators
of exterior multiplication with differential forms on loop space, or, equivalently, linear combinations
of worldsheet fermions. P denotes path ordering.

The resulting gauge covariant exterior derivative $\extd^{(A)(B)}$ on loop space will be shown to be
\begin{eqnarray}
  \label{gauge covariant loop space exterior derivative}
  \extd^{(A)(B)}
  &=&
  \extd + iT \int\limits_0^{2\pi} d\sigma\; U_A\of{2\pi,\sigma} B_{\mu\nu}X^{\prime \mu} {\cal E}^{\dagger \nu}\of{\sigma}
   U_A\of{\sigma,2\pi}
  \,,
\end{eqnarray}
where $\extd$ is the ordinary exterior derivative on loop space, $U_A$ is the holonomy of $A$ along the loop
and $U_A\of{\sigma,\kappa} = U^\dagger_A(\kappa,\sigma)$.
Note that this connection \refer{gauge covariant loop space exterior derivative}
is indeed a 1-form on loop space, taking values in the respective nonabelian
algebra carried by $B$ and $A$.

This result is similar to, but slightly and crucially different from, the construction
proposed in \cite{Hofman:2002}, the difference being the second $U_A$ factor on the right. 
It agrees however with the form of the loop space connection used in \cite{AlvarezFerreiraSanchezGuillen:1998}
in the context of integrable systems as well as with that proposed in \cite{Chepelev:2002}
in the context of non-abelian gerbes.

As a first consistency check, \emph{global} gauge transformations of \refer{gauge covariant loop space exterior derivative}
can be seen to reproduce the usual target space gauge symmetry $A \mapsto UAU^\dagger + U(dU^\dagger)$, 
$B \mapsto UBU^\dagger$.

The fact that, as we shall discuss, the operator \refer{deformation operator in introduction} also
serves as a deformation operator for boundary states, will be shown to make it quite transparent
that \emph{local} infinitesimal gauge transformations on loop space make sense only when the fermionic
terms in \refer{deformation operator in introduction} vanish. This is the case precisely when the 
1-form field strength $F_A$ cancels the 2-form field:
\begin{eqnarray}
  \label{flatness condition in introduction}
  \frac{1}{T}F_A + B = 0
  \,.
\end{eqnarray}
We will check that in this case the connection \refer{gauge covariant loop space exterior derivative}
is \emph{flat}. This is an important sufficient condition for the connection on loop space
to assign well-defined surface holonomy independent of foliation of that surface by loops.

In fact, the condition \refer{flatness condition in introduction} had been derived recently 
\cite{GirelliPfeiffer:2004} from consistency conditions in 2-group theory 
\cite{Baez:2002} (that are missing in the otherwise similar approach \cite{Chepelev:2002}). 
From the point of view of some approaches to higher gauge theory it may
seem like an obstacle for writing down interesting Lagrangians for theories. From our point of view
it seems however to be the necessary condition for a consistent coupling of the 
string to the nonabelian background. More discussion of this point is given in
\S\fullref{Flat connections on loop space and surface holonomy} and a detailed
analysis 
of nonabelian loop space connections, their relation to 2-group theory and the
conditions on non-abelian 2-forms that they imply
will be presented in \cite{Schreiber:2004g}.

In order to further clarify this point, we compute the equations of motion of the nonabelian 
background by, following \cite{Hashimoto:2000,Hashimoto:1999} (the relation to the alternative 
approach \cite{MaedaNakatsuOonishi:2004} will be discussed in \cite{Schreiber:2004d}), 
canceling divergences 
that appear when acting with \refer{deformation operator in introduction}
on the boundary state of a bare brane. The resulting equations
of motion are, to lowest order,
\begin{eqnarray}
  \mathrm{div}_A B = 0 = \mathrm{div}_A F_A
  \,,
\end{eqnarray}
where $\mathrm{div}_A$ is the gauge-covariant divergence.
In terms of the gauge field this are just the Yang-Mills equations.
Some aspects of this result will be discussed.\\

The structure of this paper is as follows:

In \S\fullref{SCFT deformations in loop space formalism} the SCFT deformation technique 
and loop space formalism studies in \cite{Schreiber:2004} is reviewed and some new aspects like
nonabelian differential forms on loop space as well as gauge connections on
loop space are discussed.

\S\fullref{BSCFT deformation for nonabelian 2-form fields} applies 
these techniques to a certain non-abelian generalization of
the previously studied abelian case and this way identifies a nonabelian connection
on loop space, coming from a 2-form on target space, as part of a superconformal algebra which should describe strings in 
nonabelian 2-form backgrounds.

Some concluding remarks are given in \S\fullref{Summary and conclusion}. 
The appendix \S\fullref{Boundary state formalism}
reviews some aspects of boundary state formalism that are referred to in the main text.
Appendix \S\fullref{computations} gives some calculation omitted from the main text.

\newpage

\newpage

\section{SCFT deformations in loop space formalism}
\label{SCFT deformations in loop space formalism}

This introductory section discusses aspects of loop space formalism and
deformation theory that will be applied in \S\fullref{BSCFT deformation for nonabelian 2-form fields}
to the description of nonabelian 2-form background fields.

\subsection{SCFT deformations and backgrounds using Morse theory technique}
\label{Deformations and backgrounds using Morse theory technique}

The reasoning by which we intend to derive the worldsheet theory for superstrings in 
nonabelian 2-form backgrounds involves an interplay of deformation theory of 
superconformal field theories for closed strings, as described in \cite{Schreiber:2004},
as well as the generalization to boundary state deformations, which are disucssed
further below in \S\fullref{Boundary state deformations from unitary loop space deformations}.
The deformation method we use consists of adding deformation terms to the super Virasoro generators
and in this respect is in the tradition of similiar approaches as for instance described in
\cite{Giannakis:1999,OvrutRama:1992,EvansOvrut:1990,EvansOvrut:1989,FreericksHalpern:1988}
(as opposed to, say, deformations of the CFT correlators).
What is new here is the systematic use of similarity transformations on a certain combination
of the supercharges, as explained below.

In this section the SCFT deformation technique for the closed string is briefly reviewed
in a manner which should alleviate the change of perspective from the string's Fock
space to loop space.

Consider some realization of the superconformal generators $L_n, \bar L_n, G_r, \bar G_r$
(we follow the standard notation of \cite{Polchinski:1998})
of the type II superstring. We are looking for consistent deformations of these operators
to operators $L^\Phi_n, \bar L^\Phi_n, G^\Phi_r, \bar G^\Phi_r$ 
($\Phi$ indicates some unspecified background field consiguration which is associated with the deformation)
which still satisfy the
superconformal algebra and so that the generator of spatial worldsheet reparametrizations
remains invariant:
\begin{eqnarray}
  \label{invariance of spatial rep}
  L^\Phi_n - \bar L^\Phi_{-n} &\shallbe& L_n - \bar L_{-n}
  \,.
\end{eqnarray}
This condition follows from a canonical analysis of the worldsheet action, which
is nothing but 1+1 dimensional supergravity coupled to various matter fields. As for all
gravitational theories, their ADM constraints break up into spatial diffeomorphism
constraints as well as the Hamiltonian constraint, which alone encodes the dynamics.

The condition \refer{invariance of spatial rep} can also be understood in terms of
boundary state formalism, which is briefly reviewed in 
\S\fullref{Boundary state formalism}. As discussed below, the operator 
$\mathcal{B}$ related to a nontrivial bounday state $\ket{\mathcal{B}}$ can be interpreted as 
inducing a deformation $G_r^\Phi \defas \mathcal{B}^{-1}G_r \mathcal{B}$, etc. and the
condition \refer{invariance of spatial rep} is then equivalent to
\refer{Virasoro constraint on boundary state}.

In any case, we are looking for isomorphisms of the superconformal algebra
which satisfy \refer{invariance of spatial rep}:

To that end,
let $d_r$ and $d^\dagger_r$ be the modes of the polar combinations of the left- and right-moving
supercurrents 
\begin{eqnarray}
  \label{polar supercurrent modes}
  d_r 
  &\defas&
    G_r + i \bar G_{-r}
  \nonumber\\
  d^\dagger_r
  \defas 
  (d_r)^\dagger 
  &=&
    G_r - i \bar G_{-r}
  \,.
\end{eqnarray}
These are the 'square roots' of the reparametrization generator
\begin{eqnarray}
  \label{rep gen}
  \mathcal{L}_n 
  &\defas&
  -i
  \left(
  L_n - \bar L_{-n}
  \right)
  \,,
\end{eqnarray}
i.e.
\begin{eqnarray}
  \antiCommutator{d_r}{d_s}
  =
  \lbrace d^\dagger_r, d^\dagger_s\rbrace
  =
  2i\mathcal{L}_{r+s}
  \,.
\end{eqnarray}
Under a deformation the right hand side of this equation must stay invariant 
\refer{invariance of spatial rep} so that
\begin{eqnarray}
  d_r^{\Phi} &\defas& d_r + \Delta_\Phi  d_r
  \nonumber\\
  d^{\dagger\Phi}_r &\defas& d^\dagger_r + (\Delta_\Phi  d_r)^\dagger
\end{eqnarray}
implies that the shift $\Delta_\Phi d_r$ of $d_r$ has to satisfy
\begin{eqnarray}
  \label{shift in loop space extd}
  \antiCommutator{d_r}{\Delta_\Phi d_s}
  +
  \antiCommutator{d_s}{\Delta_\Phi d_r}
  +
  \antiCommutator{\Delta_\Phi d_r}{\Delta_\Phi d_s}
  &=&
  0
  \,.
\end{eqnarray}

One large class of solutions of this equation is
\begin{eqnarray}
  \label{similarity transform solutions of deformation condition}
  \Delta_\Phi d_r &=& A^{-1} \commutator{d_r}{A}\,,\hspace{.8cm} 
  \mbox{for $\commutator{\mathcal{L}_n}{A} = 0\;\, \forall\, n$}
  \,,
\end{eqnarray}
where $A$ is any even graded operator that is spatially reparametrization invariant, i.e. which 
commutes with \refer{rep gen}.

When this is rewritten as
\begin{eqnarray}
  \label{simtrafo on susy currents}
  d^\Phi_r &=& A^{-1} \circ d_r \circ A
  \nonumber\\
  d_r^{\dagger \Phi} &=& A^\dagger \circ d^\dagger_r \circ A^{\dagger -1}
\end{eqnarray}
one sees explicitly that the formal structure involved here is a direct generalization of
that used in \cite{Witten:1982} in the study of the relation of deformed generators in
supersymmetric quantum \emph{mechanics} to Morse theory. Here we are concerned with 
the direct generalization of this mechanism from $1+0$ to $1+1$ dimensional 
supersymmetric field theory.

In $1+0$ dimensional SQFT (i.e. supersymmetric quantum mechanics) relation 
\refer{simtrafo on susy currents} is sufficient for the deformation to be truly an isomorphism
of the algebra of generators. In 1+1 dimensions, on the superstring's worldsheet,
there is however one further necessary condition for this to be the case.
Namely the (modes of the) new worldsheet Hamiltonian constraint 
$H_n = L_n + \bar L_-n$ 
must clearly be defined as
\begin{eqnarray}
  \label{def deformed worldsheet Hamiltonian}
  H^\Phi_n
  &\defas&
  \frac{1}{2}
  \antiCommutator{d^{\Phi}_r}{d^{\dagger\Phi}_{n-r}}
  -
  \delta_{n,0}
  \frac{c}{12}\left(4r^2 - 1\right)
\end{eqnarray}
and \refer{shift in loop space extd} alone does not guarantee that this is \emph{unique} 
for all $r \neq n/2$. 
\emph{If} it is, however, then the Jacobi identity already implies that 
\begin{eqnarray}
  G^\Phi_r &\defas& \frac{1}{2}\left(d^\Phi_r + d^{\dagger \Phi}_r\right)
  \nonumber\\
  L^\Phi_n &\defas& \frac{1}{4}\left(
    \antiCommutator{d_r^{\Phi}}{d^{\dagger \Phi}_{n-r}}
    +
    \antiCommutator{d^\Phi_r}{d^\Phi_{n-r}}
   \right)
  -
  \delta_{r,n/2}
  \frac{c}{24}\left(4r^2 - 1\right)
  \nonumber\\
  \bar G^\Phi_{r} &\defas& -\frac{i}{2}\left(d^\Phi_{-r} - d^{\dagger \Phi}_{-r}\right)
  \nonumber\\
  L^\Phi_n &\defas& \frac{1}{2}\left(
    \antiCommutator{d_{-r}^{\Phi}}{d^{\dagger \Phi}_{r-n}}
    -
    \antiCommutator{d^\Phi_{-r}}{d^\Phi_{r-n}}
   \right)  
  \,,
  \hspace{.8cm}
  \forall\,r\neq n/2
\end{eqnarray}
generate two mutually commuting copies of the super Virasoro algebra.

In order to see this first note that the two copies of the unperturbed Virasoro
algebra in terms of the 'polar' generators $d_r, d^\dagger_r, i\mathcal{L}_m, H_m$
read
\begin{eqnarray}
  &&\antiCommutator{d_r}{d_s} = 2\, i\mathcal{L}_{r+s} = \antiCommutator{d^\dagger_r}{d^\dagger_s}
  \nonumber\\
  &&
  \commutator{i\mathcal{L}_m}{d_r} = \frac{m-2r}{2}d_{m+r}
  \nonumber\\
  &&
  \commutator{i \mathcal{L}_m}{d^\dagger_r} = \frac{m-2r}{2}d^\dagger_{m+r}
  \nonumber\\
  &&
  \commutator{i\mathcal{L}_m}{i\mathcal{L}_n} = (m-n)i\mathcal{L}_{m+n}
  \nonumber\\
  &&
  \commutator{i\mathcal{L}_m}{H_n} = (m-n)i\mathcal{H}_{m+n} + \frac{c}{6}(m^3 - m)\delta_{m,-n}
  \nonumber\\
  &&
  \commutator{H_m}{d_r} = \frac{m-2r}{2}d^\dagger_{m+r}
  \nonumber\\
  &&
  \commutator{H_m}{d^\dagger_r} = \frac{m-2r}{2}d_{m+r}
  \nonumber\\
  &&
  \commutator{H_m}{H_n} = (m-n)\,i\mathcal{L}_{m+n}
  \,.
\end{eqnarray}

Now check that these relations are obeyed also by the deformed generators
$d^\Phi_r$, $d^{\dagger\Phi}_r$, $i\mathcal{L}_m$, $H^\Phi_m$
using the
two conditions \refer{simtrafo on susy currents} and \refer{def deformed worldsheet Hamiltonian}:

First of all the relations
\begin{eqnarray}
  \label{bracket Lm dr}
  \commutator{i\mathcal{L}_m}{d^\Phi_r} &=& \frac{m-2r}{2}d^\Phi_{m+r}
  \nonumber\\
  \commutator{i\mathcal{L}_m}{d^{\dagger \Phi}_r} &=& \frac{m-2r}{2}d^{\dagger \Phi}_{m+r}  
\end{eqnarray}
follow simply from \refer{simtrafo on susy currents} and the original bracket 
$\commutator{L_m}{G_r} = \frac{m-2r}{2}G_{m+r}$ and immediately imply
\begin{eqnarray}
  \commutator{i\mathcal{L}_m}{i\mathcal{L}_n} &=& (m-n)i\mathcal{L}_{m+n}
\end{eqnarray}
(note that here the anomaly of the left-moving sector cancels that of the right-moving one).

Furthermore 
\begin{eqnarray}
  \commutator{i\mathcal{L}_m}{H^\Phi_n} &=& 
  \commutator{i\mathcal{L}_m}{\frac{1}{2}\antiCommutator{d^\Phi_r}{d^{\dagger \Phi}_{n-r}}
  }
  \nonumber\\
  &\equalby{bracket Lm dr}&
  \frac{m-2r}{4}\antiCommutator{d^\Phi_{m+r}}{d^{\dagger \Phi}_{n-r}}
  +
  \frac{m-2(n-r)}{4}\antiCommutator{d^\Phi_{r}}{d^{\dagger \Phi}_{m+n-r}}
  \nonumber\\
  &\equalby{def deformed worldsheet Hamiltonian}&
  (m-n)H^\Phi_{m+n}
  +
  \delta_{m,-n}
  \frac{c}{6}
  \left(
    \frac{m-2r}{4}
    (4(m+r)^2-1)
    +
    \frac{m-2(n-r)}{4}(4r^2-1)
  \right)
  \nonumber\\
  &=&
  (m-n)H^\Phi_{m+n}
  +
  \delta_{m,-n}\frac{c}{6}\left(m^3 - m\right)
  \,.
\end{eqnarray}
(Here the anomalies from both sectors add.)

The commutator of the Hamiltonian with the supercurrents is obtained for instance by first writing:
\begin{eqnarray}
  \commutator{H^\Phi_m}{d^\Phi_r}
  &=&
  \frac{1}{2}
  \commutator{\antiCommutator{d_r^\Phi}{d^{\dagger \Phi}_{m-r}}}{d^\Phi_r}
  \nonumber\\
  &=&
  -
  \frac{1}{2}
  \commutator{
    \antiCommutator{d^\Phi_r}{d^\Phi_r}
  }
  {d^{\dagger \Phi}_{m-r}}
  -
  \frac{1}{2}
  \commutator{\antiCommutator{d_r^\Phi}{d^{\dagger \Phi}_{m-r}}}{d^\Phi_r}  
  \nonumber\\
  &=&
  -\commutator{i\mathcal{L}_{2r}}{d^{\dagger \Phi}_{m-r}}
  -
  \commutator{H_m^\Phi}{d^\Phi_r}
  \nonumber\\
   &=&
   (m-2r)d^{\dagger \Phi}_{m+r}
  -
  \commutator{H_m^\Phi}{d^\Phi_r}   
  \,,
\end{eqnarray}
from which it follows that
\begin{eqnarray}
  \commutator{H^\Phi_m}{d^\Phi_r}
  &=&
  \frac{(m-2r)}{2}d^{\dagger \Phi}_{m+r}
\end{eqnarray}
and similarly
\begin{eqnarray}
  \commutator{H^\Phi_m}{d^{\dagger \Phi}_r}
  &=&
  \frac{(m-2r)}{2}d^{\Phi}_{m+r}
  \,.
\end{eqnarray}
This can finally be used to obtain
\begin{eqnarray}
  \commutator{H^\Phi_m}{H_n^\Phi}
  &=&
  (m-n)i \mathcal{L}_{m+n}
  \,.
\end{eqnarray}

In summary this shows that every operator $A$ which 
\begin{enumerate}
\item
commutes with $i\mathcal{L}_m$
\item
  is such that
  $\antiCommutator{A^{-1} d_r A}{A^\dagger d^\dagger_{n-r} A^{\dagger -1}} - \delta_{n,0}\frac{c}{12}(4r^2-1)$
  is \emph{independent} of $r$
\end{enumerate}
defines a consistent deformation of the super Virasoro generators and hence a string background which
satisfies the classical equations of motion of string field theory.

In \cite{Schreiber:2004} it was shown how at least all massless NS and NS-NS backgrounds
can be obtained by deformations $A$ of the form $A = e^{\mathbf{W}}$,
where $\mathbf{W}$ is related to the vertex operator of the respective background field.
For instance a Kalb-Ramond $B$-field background is induced by setting
\begin{eqnarray}
  \label{B field deformation}
  \mathbf{W}^{(B)}
  &=&
  \frac{1}{2}\int d\sigma\; \left(\frac{1}{T}dA + B\right)_{\mu\nu}{\cal E}^{\dagger \mu}{\cal E}^{\dagger \nu}
  \,,
\end{eqnarray}
where ${\cal E}^\dagger$ are operators of exterior multiplication with differential forms on loop space,
to be discussed in more detail below in \S\fullref{Differential geometry on loop space}, and
we have included the well known contribution of the 1-form gauge field $A$.

Moreover, it was demonstrated in \cite{Schreiber:2003a} that the structure 
\refer{simtrafo on susy currents}
of the SCFT deformations allows to handle superstring evolution in nontrivial backgrounds
as generalized Dirac-K{\"a}hler evolution in loop space.

In the special case where $A$ is \emph{unitary} the similarity transformations 
\refer{simtrafo on susy currents} of $d$ and $d^\dagger$ 
and hence of all other elements of the super-Virasoro
algebra are identical and the deformation is nothing but a unitary transformation. It was
discussed in \cite{Schreiber:2004} that gauge transformations of the background fields,
such as reparameterizations or gauge shifts of the Kalb-Ramond field, are described by such 
unitary transformation.

In particular, an \emph{abelian} 
gauge field background was shown to be induced by the Wilson line
\begin{eqnarray}
  \label{abelian gauge field deformation}
  \mathbf{W}^{(A)} &=& i \oint d\sigma\; A_\mu\of{X\of{\sigma}}X^{\prime \mu}\of{\sigma}
\end{eqnarray}
of the gauge field along the closed string.

While the above considerations apply to closed superstrings,
in this paper we shall be concerned with open superstrings, since these carry the
Chan-Paton factors that will transform under the nonabelian group that 
we are concerned with in the context on nonablian 2-form background fields.

It turns out that the above method for obtaining closed string backgrounds by
deformations of the differential geometry of loop space nicely generalizes to
open strings when boundary state formalism is used. This is the content of the
next section.

\subsection{Boundary state deformations from unitary loop space deformations}
\label{Boundary state deformations from unitary loop space deformations}

The tree-level diagram of an open string attached to a D-brane is a disk attached to
that brane with a certain boundary condition on the disk characterizing the presence of the
D-brane. In what is essentially a generalization of the method of image charges in electrostatics 
this can be equivalently
described by the original disk ``attached'' to an auxiliary disc, so that a sphere is formed,
and with the auxiliary disk describing incoming closed strings in just such a way, that the 
correct boundary condition is reproduced. 

Some details behind this heuristic picture are recalled in
\S\fullref{Boundary state formalism}. For our purposes it suffices to note that 
a deformation \refer{simtrafo on susy currents} of the superconformal generators for
closed strings with $A$ a \emph{unitary} operator (as for instance given by 
\refer{abelian gauge field deformation}) is of course equivalent to a corresponding unitary
transformation of the closed string states. But this means that the boundary state formalism implies that 
open string dynamics in a given background described by a unitary deformation operator $A$ on loop space is
described by a boundary state $A^\dagger\ket{D9}$, where $\ket{D9}$ is the boundary state
of a bare space-filling brane, which again, as discussed below in 
\refer{condition on constant 0-form on loop space}, is nothing but the
constant 0-form on loop space.

In this way boundary state formalism rather nicely generalizes the loop space formalism 
used here from closed to open strings.

In a completely different context, the above general picture has in fact been
verified for abelian gauge fields in \cite{Hashimoto:2000,Hashimoto:1999}. 
There it is shown that acting with \refer{abelian gauge field deformation} 
and the unitary part\footnote{
  When acting on $\ket{D9}$ 
  the non-unitary part of \refer{B field deformation} is projected out automatically.
} 
of \refer{B field deformation} for $B = 0$ on $\ket{D9}$, one obtains the correct boundary state
deformation operator
\begin{eqnarray}
  \label{deform op for abelian A field}
  \exp\of{\mathbf{W}^{(A)(B = \frac{1}{T}dA)}}
  \exp\of{
    \int\limits_0^{2\pi}
   \left(
     i A_\mu X^{\prime\mu}
     +
     \frac{1}{2T}(dA)_{\mu\nu}{\cal E}^{\dagger \mu}{\cal E}^{\dagger \nu}
   \right)
  }
\end{eqnarray}
which describes open strings on a D9 brane with the given gauge field turned on.

Here, we want to show how this construction directly generalizes to deformations
describing nonabelian 1- and 2-form backgrounds. 
It turns out that the loop space perspective together with boundary state formalism
allows to identify the relation between the
nonablian 2-form background and the corresponding connection on loop space, which 
again allows to get insight into the gauge invariances of gauge theories with nonabelian 2-forms.

One simple observation of the abelian theory proves to be crucial for the non-abelian generalization:
Since \refer{deform op for abelian A field} \emph{commutes} with $\extd_K$ the loop space connection
it induces (following the reasoning to be described in \S\fullref{connections on loop space})
\emph{vanishes}. This makes good sense, since the closed string does
not feel the background $A$ field.

But the generalization of a \emph{vanishing} loop space connection to something less trivial but still trivial
enough so that it can describe something which does not couple to the closed string is a
\emph{flat} loop space connection. Flatness in loop space means that every closed curve in loop space,
which is a torus worldsheet (for the space of oriented loops) in target space, is assigned
surface holonomy $g=1$, the identity element. This means that only open worldsheets with
boundary can feel the presence of a flat loopspace connection, just as it should be.

From this heuristic picture we expect that abelian but flat loop space connections play a special
role. Indeed, we shall find in \S\fullref{2-form gauge transformations} that only these
are apparently well behaved enough to avoid a couple of well known problems.\\\

The next section 
first demonstrates that the meaning of the above constructions become
rather transparent when the superconformal generators are identified as deformed
deRham operators on loop space.

\subsection{Superconformal generators as deformed deRham operators on loop space.}

Details of the representation of the super Virasoro generators on loop space have
been given in \cite{Schreiber:2004} and we here follow the notation introduced there.

\subsubsection{Differential geometry on loop space}
\label{Differential geometry on loop space}

Again, the loop space formulation can nicely be motivated from boundary state formalism:

The boundary state $\ket{\mathrm{b}}$ 
describing the space-filling brane in Minkowski space is, according to
\refer{constraint on boundary state}, given by the constraints
\begin{eqnarray}
  \label{boundary constraints of bare D9 brane}
  \left(
    \alpha_n^\mu + \bar \alpha^\mu_{-n} 
  \right)
  \ket{\mathrm{b}}
  = 0\,, \hspace{.8cm} \forall\, n,\mu
 \nonumber\\
  \left(
    \psi_r^\mu - i \bar \psi^\mu_{-r} 
  \right)
  \ket{\mathrm{b}}
  = 0\,, \hspace{.8cm} \forall\, r,\mu
\end{eqnarray}
(in the open string R sector).

We can think of the super-Virasoro constraints
as a Dirac-K{\"a}hler system on the exterior bundle over loop space 
$\mathcal{L}\of{\manifold}$ with coordinates
\begin{eqnarray}
  \label{X mode expansion}
  X^{(\mu,\sigma)} &=& \frac{1}{\sqrt{2\pi}}X_0^\mu
  +
  \frac{i}{\sqrt{4\pi T}}
  \sum\limits_{n\neq 0}
  \frac{1}{n}
  \left(
    \alpha^\mu_n - \tilde \alpha_{-n}^\mu
  \right)
  e^{in\sigma}\,,
\end{eqnarray}
holonomic vector fields
\begin{eqnarray}
  \frac{\delta}{\delta X^\mu\of{\sigma}}
  \defas
  \partial_{(\mu,\sigma)}
  &=&
  i \sqrt{\frac{T}{4\pi}}
  \sum\limits_{n = -\infty}^\infty
  \eta_{\mu\nu}\left(
   \alpha_n^\nu
   +
   \tilde \alpha_{-n}^\nu
  \right)
  e^{in\sigma}
\end{eqnarray}
differential form creators
\begin{eqnarray}
  \label{loop space differential form}
  {\cal E}^{\dagger (\mu,\sigma)}
  &=&
  \frac{1}{2}
  \left(
    \psi^\mu_+\of{\sigma}
    + 
    i
    \psi^\mu_-\of{\sigma}
  \right)
  \nonumber\\
  &=&
  \frac{1}{\sqrt{2\pi}}
  \sum\limits_r
  \left(
    \bar \psi_{-r} + i \psi_r
  \right)
  e^{ir\sigma}
\end{eqnarray}
and annihilators
\begin{eqnarray}
  {\cal E}^{(\mu,\sigma)}
  &=&
  \frac{1}{2}
  \left(
    \psi^\mu_+\of{\sigma}
    -
    i
    \psi^\mu_-\of{\sigma}
  \right)
  \nonumber\\
  &=&
  \frac{1}{\sqrt{2\pi}}
  \sum\limits_r
  \left(
    \bar \psi_{-r} - i \psi_r
  \right)
  e^{ir\sigma}
  \,.
\end{eqnarray}
In the polar form \refer{polar supercurrent modes} the fermionic super Virasoro constraints are identified with
the modes of the exterior derivative on loop space
\begin{eqnarray}
  \label{deformed extd on loop space}
  \extd_K
  &=&
  \int\limits_0^{2\pi}
  d\sigma\;
  \left(
  {\cal E}^{\dagger \mu}
  \partial_{\mu}
  \of{\sigma}
  +
  i T X^{\prime \mu}
  {\cal E}_{\mu}
  \of{\sigma}
  \right)
  \,,
\end{eqnarray}
deformed by the reparametrization Killing vector 
\begin{eqnarray}
  \label{rep Killing vector}
  K^{(\mu,\sigma)}
  &\defas&
  X^{\prime \mu}\of{\sigma}
  \,,
\end{eqnarray}
where $T = \frac{1}{2\pi \alpha^\prime}$ is the string tension.
The Fourier modes of this operator are the polar operators of \refer{polar supercurrent modes}
\begin{eqnarray}
  d_r &\propto& \oint d\sigma\; e^{-ir\sigma} \extd_K\of{\sigma}
  \,. 
\end{eqnarray}

Using this formulation of the super-Virasoro constraints it would seem natural to 
represent them on a Hilbert space whose 'vacuum' state $\ket{\mathrm{vac}}$ is the 
\emph{constant 0-form} on loop space, i.e.
\begin{eqnarray}
  \label{condition on constant 0-form on loop space}
  \partial_{(\mu,\sigma)}\ket{\mathrm{vac}} = 0 = {\cal E}_{(\mu,\sigma)}\ket{\mathrm{vac}}
  \hspace{.8cm}
  \forall\, \mu,\sigma
  \,.
\end{eqnarray}
While this is not the usual $\mathrm{SL}\of{2,\C}$ invariant vacuum of the closed
string, it is precisely the boundary state 
\refer{boundary constraints of bare D9 brane}
\begin{eqnarray}
  \label{D9 boundary state}
  \ket{\mathrm{vac}} &=& \ket{\mathrm{b}}
\end{eqnarray}
describing the D9 brane.

For the open string NS sector the last relation of \refer{boundary constraints of bare D9 brane}
changes the sign
\begin{eqnarray}
  \left(
    \psi_r^\mu + i \bar \psi^\mu_{-r} 
  \right)
  \ket{\mathrm{b}^\prime}
  = 0\,, \hspace{.8cm} \forall\, r,\mu  \;\;\mbox{NS sector}
\end{eqnarray}
and now implies that the vacuum is, from the loop space perspective, 
the formal \emph{volume form} instead of the
constant 0-form, i.e. that form annihilated by all differential form multiplication operators:
\begin{eqnarray}
  \partial_{(\mu,\sigma)}\ket{\mathrm{b}^\prime} = 0 = {\cal E}^{\dagger (\mu,\sigma)}\ket{\mathrm{b}^\prime}
  \hspace{.8cm}
  \forall\, \mu,\sigma
  \,.
\end{eqnarray}
In finite dimensional flat manifolds of course both are related simply by \emph{Hodge duality}:
\begin{eqnarray}
  \ket{\mathrm{b}^\prime} = \star \ket{\mathrm{b}}
  \,.
\end{eqnarray}
So Hodge duality on loop space translates to the $\mbox{NS} \leftrightarrow {R}$ transition on the
open string sectors.

\paragraph{Parametrized loop space}

It is important to note that we are dealing with the space $\mathcal{L}\of{\manifold}$ 
of \emph{paramerized} loops. More precisely, for our purposes we
define this space as the closure of the space of continuous maps $X$ from the \emph{open} interval $(0,2\pi)$ 
into target space $\manifold$, such that the image is a closed loop with a single point removed:
\begin{eqnarray}
  \label{definition of loop space}
  \mathcal{L}\of{\manifold}
  &\defas&
  \overline{
  \set{
    X : (0,2\pi) \to \manifold | \lim\limits_{\epsilon \to 0} (X\of{\epsilon} - X\of{2\pi-\epsilon}) = 0
  }}
  \,.
\end{eqnarray}

This is a weak form of singling out a base point, i.e. of working on the space
of based loops, and it will turn out to be necessary in order to have a sensible notion
of reparametrization invariance in the presence of nonabelian Wilson lines along the
loops. Precisely the same phenomenon is known from all approaches to non-abelian
surface holonomy, and 
its meaning and implications will be discussed in detail 
in \S\fullref{Parallel transport to the base point} after we have derived some formulas
in the following sections.

\subsubsection{Connections on loop space}
\label{connections on loop space}

It is now straightforward to identify the relation between background fields induced by
deformations \refer{simtrafo on susy currents} and connections on loop space. A glance at
\refer{deformed extd on loop space} shows that we have to interpret the term of
differential form grade $+1$ in the polar supersymmetry generator as 
${\cal E}^{\dagger}\gradOp^{(\Phi)}_\mu$, where $\gradOp^\Phi_\mu$ is a loop space
connection (covariant derivative) induced by the target space background field $\Phi$.

Indeed, as was shown in \cite{Schreiber:2004}, one finds for instance that a gravitational background
$G_{\mu\nu}$ leads to $\gradOp^{(G)}$ which is just the Levi-Civita connection on loop
space with respect to the metric induced from target space. Furthermore, an \emph{abelian}
2-form field background is associated with a deformation operator 
\begin{eqnarray}
  \label{abelian B field deformation}
  \mathbf{W}^{(B)}
  &=&
  \oint d\sigma\
  B_{\mu\nu}{\cal E}^{\dagger \mu}{\cal E}^{\dagger \nu}
\end{eqnarray}
and leads to a connection
\begin{eqnarray}
  \label{abelian loop space connection}
  \gradOp^{(G)(B)}_\mu &=& \gradOp^{(G)}_\mu - i T B_{\mu\nu}X^{\prime \nu}
  \,,
\end{eqnarray}
just as expected for a string each of whose points carries $U(1)$ 
charge under $B$ proportional to the length element $X^\prime \,d\sigma$.

\section{BSCFT deformation for nonabelian 2-form fields}
\label{BSCFT deformation for nonabelian 2-form fields}

The above mentioned construction can now be used to examine deformations that involve
nonabelian 2-forms:

\subsection{Nonabelian Lie-algebra valued forms on loop space}

When gauge connections on loop space take values in nonabelian 
algebras deformation operators such as $\exp\of{\mathbf{W}^{(A)}}$ \refer{abelian gauge field deformation} and 
$\exp\of{\mathbf{W}^{(B)}}$
\refer{abelian B field deformation} obviously have to be replaced by path ordered exponentiated
integrals. The elementary properties of loop space differential forms involving such
path ordered integrals are easily derived, and were for instance given in 
\cite{GetzlerJonesPetrack:1991,Hofman:2002}.

So consider a differential $p+1$ form $\omega$ on \emph{target space}. It lifts
to a $p+1$-form $\Omega$ on loop space given by
\begin{eqnarray}
  \Omega
  &\defas&
  \frac{1}{(p+1)!}
  \int\limits_{S^1}
  \omega_{\mu_1\cdots \mu_{p+1}}\of{X}
   {\cal E}^{\dagger \mu_1}
  \cdots
   {\cal E}^{\dagger \mu_{p+1}} 
  \,. 
\end{eqnarray}

Let $\hat K = X^{\prime (\mu,\sigma)}{\cal E}_{(\mu,\sigma)}$ be the operator of interior
multiplication with the reparametrization Killing vector $K$ \refer{rep Killing vector}
on loop space. The above $p+1$-form
is sent to a $p$-form $\oint (\omega)$ on loop space by contracting with this Killing vector
(brackets will always denote the graded commutator):
\begin{eqnarray}
  \oint (\omega)
  &\defas&
  \commutator{\hat K}{\Omega}
  \nonumber\\
  &=&
  \frac{1}{p!}
  \int\limits_{S^1}
  d\sigma\;
  \omega_{\mu_1 \cdots \mu_{p+1}}
  X^{\prime \mu_{1}}
  {\cal E}^{\dagger \mu_2}\cdots
  {\cal E}^{\dagger \mu_{p+1}}
  \,.
\end{eqnarray}

The anticommutator of the loop space exterior derivative $\extd$ with $\hat K$
is just the reparametrization Killing Lie derivative
\begin{eqnarray}
  \commutator{\extd}{\hat K}
  &=&
  i{\cal L}_K
\end{eqnarray}
which commutes with 0-modes of fields of definite reparametrization weight, e.g.
\begin{eqnarray}
  \commutator{{\cal L}}{\Omega} &=& 0
  \,.
\end{eqnarray}
It follows that 
\begin{eqnarray}
  \label{an equation}
  \commutator{\extd }{\commutator{\hat K}{ \Omega}}
  &=&
  \commutator{\cal L}{\Omega}
  -
  \commutator{\hat K }{\commutator{\extd}{ \Omega}}
\end{eqnarray}
which implies that
\begin{eqnarray}
  \commutator{\extd}{\oint (\omega)}
  &=&
  \oint (-d\omega)
  \,.
\end{eqnarray}

The generalization to multiple path-ordered integrals
\begin{eqnarray}
  \label{path loop space integral}
  \oint (\omega_1, \cdots, \omega_n)
  &\defas&
  \int\limits_{0 < \sigma_{i-1} < \sigma_i < \sigma_{i+1} < \pi}
  \!\!\!\!\!\!\!
  d^n\sigma\;
  \commutator{\hat K}{\omega_1}\of{\sigma_1}\cdots
  \commutator{\hat K}{\omega_n}\of{\sigma_n}
\end{eqnarray}
is (see \S\fullref{The exterior derivative of path-ordered loop space forms} for the derivation)
\begin{eqnarray}
  \label{extd on path ordered forms}
  &&\commutator{\extd}{\oint (\omega_1, \cdots,\omega_n)} = 
  \nonumber\\
  &=&
  -\sum\limits_k (-1)^{\sum\limits_{i < k}p_i}
  \left(
    \oint (\omega_1, \cdots, d\omega_k, \cdots, \omega_n)
    +
    \oint (\omega_1,\cdots, \omega_{k-1}\wedge \omega_k,\cdots,\omega_n)
  \right)    
  \,.
  \nonumber\\
\end{eqnarray}
This is proposition 1.6 in \cite{GetzlerJonesPetrack:1991}.

Notice that our definition \refer{definition of loop space} 
of loop space restricts integrations over $\sigma$
to the complement of the single point $\sigma = 0 \sim 2\pi$, so that ``boundary'' terms 
$
  \omega_1\of{0}
  \oint (\omega_2,\cdots,\omega_n)
  \pm
  \oint (\omega_1,\cdots,\omega_{n-1})
  \omega_{n}\of{2\pi}
$
do not appear. 

In the light of \refer{simtrafo on susy currents} we are furthermore interested in expressions of the form
$  U_A\of{2\pi,0}
  \circ
  \extd_K
  \circ
  U_A\of{0,2\pi}$
where $U_A$ is the holonomy of $A$.

Using
\begin{eqnarray}
  \commutator{\extd}{U_A\of{0,2\pi}}
  &=&
  \commutator{\extd}{
    \sum\limits_{n=0}^\infty
    \oint \underbrace{(iA, \cdots,iA)}_{\mbox{$n$ times}}
  }
  \nonumber\\
  &=&
  -
  \sum\limits_{n=0}^\infty
  \sum\limits_{k}
  \oint 
  (iA, \cdots, iA,iF_A,iA,\cdots iA)_{\mbox{\tiny $n$ occurences of $iA$, $F_A$ at $k$}}
  \nonumber\\
  &=&
  \int\limits_0^{2\pi}
  d\sigma\;
  U_A\of{0,\sigma}
  \commutator{i F_A}{\hat K}\of{\sigma}
  U\of{\sigma,2\pi}\,
\end{eqnarray}
where
\begin{eqnarray}
  F_A &=& -i(d + iA)^2 
  \nonumber\\
  &=&
  dA + i A\wedge A
\end{eqnarray}
is the field strength of $A$ (which is taken to be hermitean),
one finds
\begin{eqnarray}
  \label{lextd deformed by Wilson line}
  U_A\of{2\pi,0}
  \circ
  \extd
  \circ
  U_A\of{0,2\pi}
  &=&
  \extd
  +
  \int\limits_0^{2\pi}
  d\sigma\,
  U_A\of{2\pi,\sigma}
  \commutator{i F_A}{\hat K}\of{\sigma}
  U_A\of{\sigma,2\pi}
  \,.
\end{eqnarray}

The point that will prove to be crucial in the following discussion is that there
is an $A$-holonomy on \emph{both} sides of the 1-form factor. The operator on the
right describes parallel transport with $A$ from $2\pi$ to $\sigma$, application of
$\commutator{\extd_A A}{\hat K}$ at $\sigma$ and then parallel transport back from $\sigma$
to $2\pi$. Following \cite{Hofman:2002} the abbreviating notation
\begin{eqnarray}
  \label{def ointA}
  \oint_A (\omega)
  &\defas&
  \int_0^{2\pi}
  d\sigma\;
  U_A\of{2\pi,\sigma}
  \commutator{\hat K}{\omega}
  U_A\of{\sigma,2\pi}
\end{eqnarray}
will prove convenient. (But notice that in \refer{def ointA} there is also a factor 
$U_A\of{\sigma,2\pi}$ on the  \emph{right}, which does not appear in \cite{Hofman:2002}.)
Using this notation \refer{lextd deformed by Wilson line} is rewritten as
\begin{eqnarray}
  \label{pure Wilson line deformation}
  U_A\of{2\pi,0}
  \circ
  \extd
  \circ
  U_A\of{0,2\pi}
  &=&
  \extd -i \oint_A (F_A)
  \,.  
\end{eqnarray}

This expression will prove to play a key role in the further development. In order
to see why this is the case we now turn to the computation and disucssion of the
connection on loop space which is induced by the nonabelian 2-form background.

\subsection{Nonabelian 2-form field deformation}
\label{Nonabelian 2-form field deformation}

With the above considerations it is now immediate how to incorporate a 
nonabelian 2-form in the target space of a boundary superconformal field
theory on the worldsheet. The direct generalization of 
\refer{abelian gauge field deformation} and 
\refer{abelian B field deformation} is obviously the deformation operator
\begin{eqnarray}
  \label{nonab B deformation}
  \exp\of{\bf W}^{(A)(B)_\mathrm{nonab}}
  &=&
  \mathrm{P}
  \exp\of{\int\limits_0^{2\pi}
    d\sigma\;
    \left(
      i A_\mu X^{\prime\mu}
      +
      \frac{1}{2}
      \left(\frac{1}{T}F_A + B\right)_{\mu\nu} {\cal E}^{\dagger \mu}{\cal E}^{\dagger \nu}
    \right)
  }
\end{eqnarray}
for non-abelian and hermitean $A$ and $B$. (P denotes path ordering)
Note that this is indeed reparametrization invariant on $\mathcal{L}\of{\manifold}$
\refer{definition of loop space} and that
there is no trace in \refer{nonab B deformation}, so that
this must act on an appropriate bundle\footnote{In this paper however only
local aspects of such a bundle play a role.}, 
which is naturally associated with a stack of $N$ branes (\cf pp. 3-4 of \cite{Zunger:2002}).

According to \S\fullref{connections on loop space} the loop space connection induced by
this deformation operator is given by the term of degree $+1$ in the deformation
of the superconformal generator \refer{deformed extd on loop space}. Using
\refer{lextd deformed by Wilson line} this is found to be
\begin{eqnarray}
  \label{non-abelian 2-form field deformation}
  \exp\of{-{\bf W}}^{(A)(B)_\mathrm{nonab}}
    \circ
    \extd_K
    \circ
  \exp\of{\bf W}^{(A)(B)_\mathrm{nonab}}
  &=&
  \extd
  +iT
  \oint_A (B)
  \nonumber\\
  &&\;\;
  +
  \mbox{(terms of grade $\neq 1$)}
  \,,
  \nonumber\\
\end{eqnarray}
where the notation \refer{def ointA} is used.

The second term $iT\oint_A (B)$ is the nonablian 1-form connection on loop space which is induced
by the target space 2-form $B$. Note that the terms involving
the $A$-field strength $d_A A$ coming from the $X^\prime$ term and 
those coming from the ${\cal E}^\dagger{\cal E}^\dagger$ 
term in \refer{non-abelian 2-form field deformation} mutually cancel. 

The connection \refer{non-abelian 2-form field deformation} is essentially that given in 
\cite{Hofman:2002}, instead that here the $U_A$ holonomy acts from both sides,
as in the expressions given in \cite{AlvarezFerreiraSanchezGuillen:1998,Chepelev:2002}. 
It will be shown below that this is crucial for the correct gauge invariance on target space.
In \cite{Schreiber:2004g} it will be discussed in detail how this very form of the gauge
connection makes the non-abelian 2-form formalism derived here from SCFT
deformations equivalent to 
that of 2-group theory \cite{Baez:2002,GirelliPfeiffer:2004}.

There is a simple way to understand the form of \refer{non-abelian 2-form field deformation}
heuristically: A point particle couples to a 1-form, a string to a 2-form. Imagine the
worldsheet foliated into spacelike hyperslices. Each \emph{point} on these is similar
to a point particle coupled to that 1-form which is obtained by contracting the 2-form with the
tangent vector to the slice at that point. In the abelian case all these contributions
can simply be summed up and so one obtains a corresponding 1-form on loop space, namely
\refer{abelian loop space connection}. But really, as noted in \cite{Hofman:2002} in general 
one has to be more careful, since elements of fibers at different points can not be compared. Instead,
we can relate the 1-forms at each point of the string by parallel tranporting them with respect to
some 1-form connection $A$ to a given (arbitrary) origin. This is precisely what is accomplished by 
the $U_A$ factors in \refer{non-abelian 2-form field deformation}.

It should be emphasized that even though the construction \refer{non-abelian 2-form field deformation}
involves a unitary transformation
of the superconformal generators this does \emph{not} imply that the 2-form gauge connection
\refer{non-abelian 2-form field deformation} is trivial up to gauge. It is crucial that the
deformation of the term $iT {\cal E}_\mu X^{\prime \mu}$ in $\extd_K$ 
\refer{deformed extd on loop space} also contributes to the gauge connection. This way, the result
is \emph{not} a unitary transform of the loop space exterior derivative $\extd$, but only
of its $K$-deformed relative $\extd_K$. Of course precisely this effect could already be
observed in simpler examples of abelian backgrounds discussed in \cite{Schreiber:2004}.

Still, we will see below that precisely the case where there is \emph{no} contribution
from $iT {\cal E}_\mu X^{\prime \mu}$ will turn out to be the most interesting one.\\

Before discussing gauge transformations and equations of motion it is in order
to have a careful look at a certain technicality:

\subsection{Parallel transport to the base point}
\label{Parallel transport to the base point}

One point deserves further attention: Note that in \refer{nonab B deformation}
we did \emph{not} include a trace over the path-ordered exponential. The reason for that is quite simple:
If we included a trace, making the formerly group-(representation)-valued expression a scalar, this
could not give rise to an exterior derivative on loop space which is covariant with respect to the given
non-abelian gauge group, simply because its connection term would be locally a loop space 1-form taking
scalar values instead of non-abelian Lie-algebra values. This way the relation to non-abelian 
surface holonomy would be completely lost and one could not expect the associated SCFT to
describe any non-abelianness of the background.

There is a further indication that not taking the trace 
is the correct thing to do: Recall from the discussion at the beginning of this section
(\cf table \ref{dimensional ladder}) that we should expect states of strings in a non-abelian
2-form background to have a single Chan-Paton-like degee of freedom, i.e. to carry one fundamental index
of the gauge group. This implies that in particular boundary states carry such an index and hence
any deformation operator acting on them must accordingly act on that index. This is precisely
what an un-traced generalized Wilson line as in \refer{nonab B deformation} does.

However, by not taking the trace in \refer{nonab B deformation}
the point $\sigma= 0 \sim 2\pi$ on the string/loop becomes a preferred point in a sense. The object 
\refer{nonab B deformation} is reparameterization invariant only under those
reparameterizations that leave the point $\sigma= 0\sim 2\pi$ where it is. Commuting it 
with a general generator of $\sigma$-reparameterizations produces two ``boundary'' terms 
(even though we do not really have any boundary at $\sigma = 0\sim 2\pi$ on the closed string)
at  $\sigma= 0$ and $\sigma = 2\pi$ which only mutually cancel when traced over.

This might at first sight appear as a problem for our proposed way to study worldsheet theories
in non-abelian 2-form backgrounds. After all, reparameterization invariance on the string
must be preserved by any reasonable physical theory. But a closer look at the general theory
of non-abelian surface holonomy indicates how the situation should clarify:

As soon as a non-abelian 2-form is considered and any notion of non-abelian surface-holonomy
associated with it, a fundamental question that arises is at which \emph{point} the 
non-abelian holonomy of a given surface ``lives'' \cite{Hofman:2002}. 
It must be associated with some
point, because it really lives in a given fiber of a gauge bundle over target space, and it has to be
specified in which one. This issue becomes quite rigorously clarified in the 
2-group description of surface holonomy 
\cite{GirelliPfeiffer:2004,Pfeiffer:2003,Baez:2002}, whose basic mechanism is
summarized from a physical point of view in \cite{Schreiber:2004g}. In the
2-group description non-abelian surface holonomy is a functor that assigns
2-group elements to ``bigons'', which are surfaces with two special
points on their boundary
(a ``source'' and a ``target'' vertex). It is a theorem 
(e.g. proposition 5 in \cite{Baez:2002})
that every 2-group
uniquely comes from a ``crossed module'' (a tuple of two groups, one associated with
our 1-form $A$ the other with the 2-form $B$ together with a way for $A$ to act on $B$
by derivations) in such a way that when composing any two bigons, their total surface
holonomy is obtained by parallel transporting (with respect to the 1-form A) 
group elements from the individual ``source'' vertices to the common source vertex
of the combined bigon. This theorem shows that it is inevitable to associate
surface holonomy with ``preferred'' points. 

On the other hand, these points are not absolutely preferred. One can choose any other point on the
boundary of the given bigon as a the source vertex. This is done by, again, parallel transporting
the surface holonomy from the original source vertex to the new one.
A more detailed description of this and related facts of 2-group theory
is beyond the scope of this paper, but can be found in 
\cite{Schreiber:2004g}, which relates 2-group theory as described in 
\cite{GirelliPfeiffer:2004,Pfeiffer:2003,Baez:2002} to the loop space formalism used here
and in \cite{AlvarezFerreiraSanchezGuillen:1998}.
Indeed, it is shown there that the preferred point $\sigma = 0\sim 2\pi$ of
\refer{nonab B deformation} is to be identified with the 
source vertex in 2-group theory.

That this is true can indeed be seen quite clearly from the expression
\refer{non-abelian 2-form field deformation} given below, which we find for the 
non-abelian covariant loop space exterior derivative (\cf with \refer{def ointA}
for the notation used there)
obtained by deforming the ordinary loop space exterior derivative
with \refer{nonab B deformation} (as described in detail
below).
 Indeed, the loop space covariant
exterior derivative to be discussed in the following involves integrating
the non-abelian 2-form B over the string, while parallel transporting its
value back from every point to $\sigma = 2\pi$. As mentioned above, and as already noted
in \cite{Hofman:2002}, this parallel transport back to a common reference
point is necessary in order to obtain a  well defined element of a fiber in 
a non-abelian gauge bundle.

It should be plausible from these comments,
and  is proven in detail in \cite{Schreiber:2004g},  that, given
non-abelian differential forms $A$ and $B$,
the surface holonomy computed using the covariant loop space connections
discussed here coincide with those computed by 2-group theory, when 
the former are integrated over loops in \emph{based} loop space, i.e. the
space of all loops with a common point in target space.

But this should finally clarify the appearance of the preferred point $\sigma = 0$
in \refer{nonab B deformation} also from the point of
view of worldsheet reparameterization invariance: Whenever we work with a non-abelian
2-form, the worldsheet surpercharges, which are generalized exterior derivatives
on loop space \cite{Schreiber:2004}, must be regarded as operators on \emph{based}
loop space (for some given base point $x$), in terms of which, for the reasons just discussed,
all computations must necessarily take place. However, the choice of this base point
is arbitrary, as surface holonomies computed with differing base points are
related simply by parallel transport between the two base points.

The restriction to based loop space may seem like a drastic step from the string
worldsheet point of view, even though the above general considerations make it
seem to be quite inevitable. On the other hand, the boundary state formalism that
we will be concerned with in the following involves spherical closed string worldsheets
at tree level, and these are precisely the closed curves in based loop space
(\cf \cite{AlvarezFerreiraSanchezGuillen:1998}).

Apart from loop space and 2-group methods a further conceptual framework for 
surface holonomy is the theory of \emph{gerbes} \cite{Chatterjee:1998}. 
Gerbes have been applied with much avail to string theory in the presence of 2-form
fields, but so far mostly in the abelian case. Approaches to construct a theory
of non-abelian gerbes can be found in \cite{Kalkkinen:1999,AschieriCantiniJurco:2004,AschieriJurco:2004}, 
but, to the best of our knowledge, it is not yet understood how to compute non-abelian
surface holonomy using gerbes. One proposal is \cite{Chepelev:2002} which makes use of
precisely the parallel transport to common base points known from 2-group theory and
found from the SCFT deformations used here (but without taking consistency conditions on the
uniqueness of surfacde holonomy into account).
On the other hand, the notion of abelian surface holonomy from abelian gerbes is well understood 
\cite{Chatterjee:1998,CareyJohnsonMurray:2004}. 
It can be expressed \cite{MackaayPicken:2001} in precisely
the loop space language used here where it, too, works on based loop space.

Even with so many compelling formal reasons to have a preferred point on the string
in the presence of non-abelian 2-form background it would be nice to also have a
more physically motivated interpretation of this phenomenon. One aspect that
might be relavant is the observation in \cite{Witten:1997b}
that at least in certain 6-dimensional theories certain strings can be regarded as
coming from M-theory membranes
that stretch between points identified under a $Z_r$ orbifold action. A similar
mechanism was discussed in \cite{SchwarzSen:1995} in the context of compactifications
of type IIA strings down to six dimensions. Maybe the preferred worldsheet point $\sigma = 0$
that we found above to be necessary for the coupling to the non-abelian 2-form field
has to be identified as attached to one of the points identified under the orbifold action.

Given all this, the conceptual issues that we encounter in our worldsheet perspective
approach to non-abelian 2-form field backgrounds  are somewhat unfamiliar, but are quite in 
accordance with both the general features expected of the target space theory as well as
general facts known about non-abelian 2-gauge theory. While it can't completely solve the
issue of superstring propagation in non-abelian 2-form backgrounds, the present approach
is hoped to illuminate some crucial aspects of any complete theory that does so, and
in particular of its worldsheet formulation.\\

With that technical point discussed we now turn to
the issue of gauge transformations of the loop space connection \refer{non-abelian 2-form field deformation}
and how this relates to gauge transformations on target space.

\subsection{2-form gauge transformations}
\label{2-form gauge transformations}

In a gauge theory with a nonabelian 2-form one expects the usual gauge invariance
\begin{eqnarray}
  \label{1st gauge trafo}
   A &\mapsto& U\,A\,U^\dagger + U(dU^\dagger)
  \nonumber\\
  B &\mapsto& U\,B\,U^\dagger
\end{eqnarray}
together with some nonabelian analogue of the infinitesimal shift
\begin{eqnarray}
  \label{2nd gauge trafo}
  A &\mapsto& A + \Lambda
  \nonumber\\
  B &\mapsto& B - d_A \Lambda + \cdots
\end{eqnarray}
familiar from the abelian theory.

With the above results, it should be possible to derive some properties of the gauge invariances of 
a nonabelian 2-form theory from loop space reasoning. That's because on loop space 
\refer{non-abelian 2-form field deformation} is an ordinary
\emph{1-form} connection. The ordinary 1-form gauge transformations of that loop space connection should 
give rise to something like \refer{1st gauge trafo} and \refer{2nd gauge trafo} automatically.

Indeed, \emph{global} gauge transformations of 
\refer{non-abelian 2-form field deformation} on loop space give rise to 
\refer{1st gauge trafo}, while infinitesimal gauge transformations on loop space
give rise to \refer{2nd gauge trafo}, but with correction terms that only have
an interpretation on loop space.

More precisely, let $U\of{X} = U$ be any \emph{constant} group valued function on (a local patch of) loop space
and let $V(X) : \manifold \to G$ 
be such that $\lim\limits_{\epsilon \to 0} V(X)\of{X\of{\epsilon}} = V(X)\of{X\of{2\pi - \epsilon}} = U$, then
(see \S\fullref{computations} for the details)
\begin{eqnarray}
  \label{1st gauge trafor from loop space}
  U \left(\oint_A (B)\right) U^\dagger + U(dU^\dagger)
  &=&
  \oint_{A^\prime}(B^\prime)
\end{eqnarray}
with
\begin{eqnarray}
  A^\prime &=& V\, A \, V^\dagger + V(dV^\dagger)
  \nonumber\\
  B^\prime &=& V\, B\, V^\dagger
  \,,
\end{eqnarray}
which reproduces \refer{1st gauge trafo}.

If, on the other hand, $U$ is taken to be a \emph{nonconstant} infinitesimal gauge transformation 
with a 1-form gauge parameter $\Lambda$ of the form
\begin{eqnarray}
  U\of{X} &=& 1 - i\oint_A (\Lambda)
\end{eqnarray}
then
\begin{eqnarray}
  U \left(\oint_A (B)\right)U^\dagger + U(dU^\dagger)
  &=&
  \oint_{A + \Lambda} (B + d_A B)
  +
  \cdots
  \,.
\end{eqnarray}

The first term reproduces \refer{2nd gauge trafo}, but there are further terms which 
do not have analogs on target space.

The reason for this problem can be understood from the boundary deformation operator point of view:

Let 
\begin{eqnarray}
  \mathrm{P} \exp\of{i\int\limits_0^{2\pi} R}
  =
  \lim\limits_{N=1/i\epsilon \to \infty} (1 + i\epsilon R(0))\cdots (1+i\epsilon R\of{\epsilon 2\pi})\cdots
  (1 + i\epsilon R\of{2\pi})
\end{eqnarray}
be the path ordered integral over some object $R$. Then a small shift $R \to R + \delta R$ amounts to
the ``gauge transformation''
\begin{eqnarray}
  \label{shift in boundary deformation operator}
  \mathrm{P} \exp\of{i\int\limits_0^{2\pi} R}
  &\to&
  \mathrm{P} \exp\of{i\int\limits_0^{2\pi} R}
  \left(
    1 + i\oint_R \delta R
  \right)
\end{eqnarray}
to first order in $\delta R$. 
Notice how, using the definition of $\oint_R$ given in \refer{def ointA}, the term $\oint_R \delta R$
inserts $\delta R$ successively at all $\sigma$ in the preceding Wilson line.

So it would seem that $U = 1 - i \oint_R \delta R$ is the
correct unitary operator, to that order, for the associated transformation. But the
problem ist that $R$ is in general not purely bosonic, but contains fermionic
contributions. These spoil the ordinary interpretation of the above $U$ as a gauge
transformation.

This means that an ordinary notion of gauge transormation is obtained if and only
if the fermionic contributions in 
\refer{nonab B deformation} disappear, which is the case when 
\begin{eqnarray}
  \label{flatness condition}
  B &=& - \frac{1}{T}F_A
  \,.
\end{eqnarray}
Then 
\begin{eqnarray}
  \label{pure A Wilson line}
  \exp\of{\mathbf{W}}^{(A)(B = - \frac{1}{T}F_A)}
  &=&
  \mathrm{P}
  \exp\of{\int\limits_0^{2\pi} d\sigma\; i A_\mu X^{\prime\mu}\of{\sigma}}
\end{eqnarray}
is the pure $A$-Wilson line
and and the corresponding gauge covariant exterior derivative on loop space is
\begin{eqnarray}
  \label{flat loop space connection}
  \extd^{(A)(B)} &=& \extd - i \oint_A (F_A)
  \,,
\end{eqnarray}
as in \refer{pure Wilson line deformation}.
Now the transformation 
\begin{eqnarray}
  A &\mapsto& A + \Lambda
  \nonumber\\
  B &\mapsto& B - \frac{1}{T}\left( d\Lambda + i A \wedge \Lambda + i \Lambda \wedge A\right)
\end{eqnarray}
is correctly, to first order, induced by the loop space gauge transformation
\begin{eqnarray}
  U\of{X} &=& 1 - i \oint_A (\Lambda)
  \,.
\end{eqnarray}
One can check explicitly that indeed
\begin{eqnarray}
  U \circ (\extd -i \oint_A (F_A)) \circ U^\dagger
  &=&
  \extd
  -
  i
  \oint_A (F_A + d\Lambda + i A\wedge \Lambda + i \Lambda \wedge A)
  + \order{\Lambda^2}
  \nonumber\\
  &&
  -
  \int\limits_{\sigma_1 \geq \sigma_2} 
  U_A\of{2\pi,\sigma_1} (-iF_A)\of{\sigma_1} U_A\of{\sigma_1,\sigma_2} i\Lambda_\mu X^{\prime \mu}\of{\sigma_2}
  U_A\of{\sigma_2,2\pi}
  \nonumber\\
  &&+
  \int\limits_{\sigma_2 \geq \sigma_1} 
  U_A\of{2\pi,\sigma_1} i\Lambda_\mu X^{\prime \mu}\of{\sigma_1}
  U_A\of{\sigma_1,\sigma_2}
  (-iF_A)\of{\sigma_2}
  U_A\of{\sigma_2,2\pi}
   \nonumber\\
  &&+
  \oint_A (-iF_A) \oint_A (i\Lambda)
  +
  \oint_A (-i\Lambda) \oint_A (-iF_A)
  \nonumber\\
  &&
  + \order{\Lambda^2}
  \nonumber\\
  &=&
  \extd -i \oint_{A + \Lambda} (F_{A + \Lambda}) + \order{\Lambda^2}
  \,,
\end{eqnarray}
as it must be.

One can furthermore check that the loop space curvature of the connection \refer{flat loop space connection}
\emph{vanishes},
as follows also directly from \refer{pure Wilson line deformation}:
\begin{eqnarray}
  \label{flatness of loopspace connection}
  \left(
    \extd - i \oint_A (F_A)
  \right)^2
  &=& 
  0
  \,.
\end{eqnarray}

A flat connection on loop space implies that the surface holonomy associated with 
tori in target space is trivial. 
As discussed at the end of \S\fullref{Boundary state deformations from unitary loop space deformations}, 
this makes sense, since the nonabelian 
connection should not couple to the closed string without boundary, due to lack of Chan-Paton factors or anything that
could play their role. 

The above shows that for flat loop space connections the expected gauge invariances
\refer{1st gauge trafo} and \refer{2nd gauge trafo} of 2-form gauge theory do hold 
without problematic correction terms and 
that apparently a consistent physical picture is obtained.

The same result has been obtained recently in \cite{GirelliPfeiffer:2004} using the
theory of 2-groups as introduced in \cite{Baez:2002}.

\subsection{Flat connections on loop space and surface holonomy}
\label{Flat connections on loop space and surface holonomy}

In the light of the flatness \refer{flatness of loopspace connection} of the connection
\refer{flat loop space connection}
in this section some general aspects of flat connections on loop space  and their relation
to parameterization invariant surface holonomies are discussed in the following.\\

Denote by $\mathcal{L}\mathcal{L}\of{\mathcal{M}}$ the space of parameterized 
loops in loop space. The holonomy
of the loop space connection $\oint_A (B)$ around these loops in loop space is a map
\begin{eqnarray}
  H : \mathcal{L}\mathcal{L}\of{\mathcal{M}} \to G
  \,,
\end{eqnarray}
where $G$ is the gauge group. This computes the \emph{surface holonomy} of the (possibly degenerate) 
unbounded surface in target space associated with a given loop-loop in $\mathcal{L}\mathcal{L}\of{\mathcal{M}}$.

In general,
there are many points in $\mathcal{L}\mathcal{L}\of{\mathcal{M}}$ that map in a bijective way to the same 
surface in $\mathcal{M}$ and that are related by reparameterization. 
Only if the function $H$ takes the same value on all these points does
the loop space connection $\oint_A (B)$ induce a well-defined surface holonomy in target space.

In the case considered above, where, locally at least, $\mathcal{M} = \R^D$, all loops in 
$\mathcal{L}\of{\mathcal{M}}$ are contractible. This means that when $\oint_A (B)$ is 
flat, $H$ maps \emph{all of} $\mathcal{L}\mathcal{L}\of{\mathcal{M}}$ to the identity element in $G$. In this
case the surface holonomy is therefore trivially well defined, since all closed surfaces represented
by points on $\mathcal{L}\mathcal{L}\of{\mathcal{M}}$ are assigned the same surface holonomy,
$H = 1$.

This is the case that has been found here to arise from boundary state deformations. It implies
that unbounded closed string worldsheets do not see the nonabelian background. But worldsheets
having a boundary that come from cutting open those surfaces corresponding to points in
$\mathcal{L}\mathcal{L}\of{\mathcal{M}}$ do. Is their surface holonomy also well defined?

For the special connection \refer{flat loop space connection} it is. This follows from the 
fact that this is just the trivial connection, which assigns the unit to everything,
gauge transformed with \refer{flat loop space connection}.
Under a gauge transformation on loop space the surface holonomies of bounded surfaces coming from 
open curves in $\mathcal{L}\of{\mathcal{M}}$ are simply multiplied from the left 
by the gauge transformation function \refer{flat loop space connection} evaluated at one boundary
$B_1$
and from the right by its inverse evaluated at the other boundary $B_2$:
\begin{eqnarray}
  H_\mathrm{bounded} = \exp\of{\mathbf{W}}\of{B_1}\exp\of{\mathbf{W}}^{-1}\of{B_2}
  \,.
\end{eqnarray}
But $\exp\of{\mathbf{W}}$
is reparameterization invariant on the loop, as long as the preferred point $\sigma = 0$ is 
unaffected, at which the Wilson loop is open (untraced). Therefore the surface holonomy induced by 
\refer{flat loop space connection} on open
curves in $\mathcal{L}\of{M}$ depends (only) on the position of the preferred point $\sigma = 0$.

In the
boundary state formalism, the point $\sigma = 0$ has to be identified with the insertion of the 
open string state which propagates on half the closed string worldsheet. So this dependence 
appears to make sense.

In the case when $\pi_2\of{\mathcal{M}}$ is nontrivial it is not as obvious to decide if
a flat connection on loop space associates unique surface holonomy with surfaces in $\mathcal{M}$.

A sufficient condition for this to be true is that
any two points in $\mathcal{L}\mathcal{L}\of{\mathcal{M}}$ which map to the same given surface
can be connected by continuously deforming the corresponding loop on $\mathcal{L}\of{\mathcal{M}}$.
This is not true for nondegenerate toroidal surfaces in $\mathcal{M}$. But it is the case for
spherical surfaces, for which the loop on $\mathcal{L}\of{\mathcal{M}}$ must begin and end
at an infinitesimal loop. All slicings of the sphere are
continuously deformable into each other, corresponding to a deformation of a loop on
$\mathcal{L}\of{\mathcal{M}}$, under which the holonomy of a flat connection is invariant.

Therefore even with nontrivial $\pi_2\of{\mathcal{M}}$ a flat connection on loop space
induces a well defined surface holonomy on topologically spherical surfaces in $\mathcal{M}$.

Whether the same statement remains true for toroidal surfaces is not obvious. But actually
for open string amplitudes at tree level in the boundary state formalism spherical surfaces
are all that is needed.\\

Due to these considerations it is an interesting task to try to characterize all flat connections
on parameterized loop space. The connection \refer{flat loop space connection} is obatined
by loop space gauge transformations from the trivial connection. Are there other flat connections?

Some flat connections on loop space were investigated in the context of integrable systems in 
\cite{AlvarezFerreiraSanchezGuillen:1998}. The authors of that paper used the same general form
\refer{non-abelian 2-form field deformation} of the connection on loop space 
that dropped out from deformation theory in our approach. They then demanded that the $A$-curvature
vanishes, $F_A \shallbe 0$, and checked that for this special case
an $A$-covariantly constant $B$, as well as an $A$-closed $B$ which furthermore
takes values in an abelian ideal, is sufficient to flatten the connection $\oint_A (B)$.
In the notation used here, this can bee seen as follows:

The curvature of the general connection \refer{non-abelian 2-form field deformation} is 
\begin{eqnarray}
  \mathbf{F}^{(A)(B)}
  &\defas&
  -i
  \left(\mathbf{d}^{(A)(B)}\right)^2
  \nonumber\\
  &=&
  -i
  \left(
    \mathbf{d} + iT \oint_A (B)
  \right)^2
  \nonumber\\
  &=&
  T \oint_A (d_A B)
  + i T^2 \oint_A (B) \; \oint_A (B)
  +
  (\mbox{terms proportional to $F_A$})
  \,.
\end{eqnarray}
For vanishing $A$-field strength this reduces to
\begin{eqnarray}
  \mathbf{F}^{(A)(B)}
  &\stackrel{F_A = 0}{=}&  
  T \oint_A (d_A B)
  + i T^2 \oint_A (B) \; \oint_A (B)  
  \,.
\end{eqnarray}
It is easy to see under which conditions both terms on the right hand side vanish by themselves  
(while it seems hardly conceivable that there are conditions under which these two terms cancel
mutually without each vanishing by themselves), namely 
the first term vanishes when $d_A B  = 0$, while the second term vanishes when the components of the
1-form $\oint_A (B)$ mutually commute.

This is for instance the case when $B$ takes values in an abelian ideal or if 
$B$ is $A$-covariantly constant and all components of $B$ at a given point commute.
These are the two conditions discussed in \S3.2 of \cite{AlvarezFerreiraSanchezGuillen:1998}.

Since these two cases correspond to abelian $B$ we have to think of them as special cases of the
theory of abelian 2-form fields. To obtain a scenario with nonabelian $B$ from these one can
again, as we did for the trivial connection, make a gauge transformation \refer{flat loop space connection} 
on loop space with respect to the holonomy of yet another 1-form connection, not nessecarily a flat one. This
way one obtains two further classes of flat connections on loop space.

Finally it should be noted that we haven't ruled out the possibility that there are non-flat connections
on loop space which induce a well defined surface holonomy. (Of course for the abelian case 
all connections of the form \refer{abelian loop space connection} induce well defined surface holonomy.)
Our main point in \S\fullref{Nonabelian 2-form field deformation} was merely that boundary state deformation
theory leads to supersymmetry generators which incorporate a flat connection.

A more detailed analysis of the consistency condition on loop space connections
to produce a well defined surface holonomy will be given in
\cite{Schreiber:2004g}.

$\,$\\Before discussing this further
in the concluding section it pays to first see what the worldsheet theory has to say
about the equations of motion of the nonabelian background field. This is the content
of the next section.

\subsection{Background equations of motion}

It was demonstrated in \cite{Hashimoto:2000,Hashimoto:1999} that
the conditions under which the operator \refer{deform op for abelian A field} for \emph{abelian} $A$,
\emph{without} any normal ordering, is well defined when acting on the boundary state
$\ket{\mathrm{b}}$ \refer{D9 boundary state} describing a bare D9 brane, is equivalent to 
the background equations of motion of $A$, at least up to second order.

A similar statement should be true for the nonabelian generalization \refer{nonab B deformation}
that we are concerned with here. Due to the consistency condition 
\refer{flatness condition} we need only consider the
case where the worldsheet fermions $\mathcal{E}^\dagger$ are all canceled by the presence of the $B$-field,
so that the background equations of motion for the nonabelian $B$-field should arise as the condition for
the canceling of divergences in the application of \refer{pure A Wilson line} to $\ket{\mathrm{b}}$.

This task is greatly simplified by working with gauge connections $A$ which
are \emph{constant} on spacetime. Due to standard arguments (e.g. \cite{AokiIsoKawaiYoshihisaKitazawaTsuchiyaTada:1999})
this should be no restriction of generality
but it greatly simplifies the computation of divergences, since the number of contractions of worldsheet
fields is very much reduced.

Namely in the case of constant $A$, i.e. $\partial_\mu A_\nu = 0$, the only divergences come
from terms of the form $X^{\prime \mu}(\kappa)X^{\prime \mu}(\sigma)\ket{\mathrm{b}}$. 
Using the mode expansion 
\refer{X mode expansion} 
as well as the boundary condition 
\refer{boundary constraints of bare D9 brane}
this contraction is seen to produce
\begin{eqnarray}
  X^{\prime \mu}(\kappa)X^{\prime \nu}(\sigma)\ket{\mathrm{b}}
  &=&
  \alpha^\prime
  \eta^{\mu\nu}
  \sum\limits_{n > 0}
  n \cos\of{n(\sigma-\kappa)}
  \ket{\mathrm{b}}
  +
  :X^{\prime \mu}(\kappa)X^{\prime \nu}(\sigma):\ket{\mathrm{b}}  
  \,,
\end{eqnarray}
where $\alpha^\prime = 1/2\pi T$.

When this expression is inserted into the expansion of \refer{pure A Wilson line}:
\begin{eqnarray}
  \label{deformation expansion}
  \mathrm{P}\exp\of{i\int\limits_0^{2\pi} d\sigma\; A_\mu X^{\prime\mu}\of{\sigma}}
  \ket{\mathrm{b}}
  &=&
  \ket{\mathrm{b}} + i A_\mu \underbrace{\int\limits_0^{2\pi} d\sigma\;X^{\prime \mu}\of{\sigma}}_{=0}\ket{\mathrm{b}}
  \nonumber\\
  &&
  -
  A_\mu
  A_\nu
  \int\limits_{0 < \sigma_1 < \sigma_2 < 2\pi}
  d\sigma_1\,d\sigma_2\;
  X^{\prime \mu}\of{\sigma_1}X^{\prime \nu}\of{\sigma_2}\ket{\mathrm{b}}
  \nonumber\\
  &&
  -i
  A_\mu A_\nu A_\lambda
  \int\limits_{0 < \sigma_1 < \sigma_2 < \sigma_3 < 2\pi}
  d\sigma_1\,d\sigma_2\,d\sigma_3 \;
  X^{\prime \mu}\of{\sigma_1}X^{\prime \nu}\of{\sigma_2}X^{\prime \lambda}\of{\sigma_3}\ket{\mathrm{b}}  
  \nonumber\\
  && + \cdots
\end{eqnarray} 
one immediately sees that the $A^2$ term does not produce any divergence so that the first nontrivial
case is the $A^3$ term. A simple calculation yields
\begin{eqnarray}
    \int\limits_{0 < \sigma_1 < \sigma_2 < \kappa}\!\!\!\!\!\!\!\!\!\!\! d\sigma_1\, d\sigma_2\;
    \cos\of{n(\sigma_1-\sigma_2)}  
    =
    \!\!\!\!\!\!\!\!\!\!\!\!
    \int\limits_{\kappa < \sigma_1 < \sigma_2 < 2\pi}\!\!\!\! \!\!\!\!\!\!\!d\sigma_1\, d\sigma_2\;
    \cos\of{n(\sigma_1-\sigma_2)}
    =
    -\frac{1}{2}\!\!\!\!\!\!\!\!\!\!\!\!
    \int\limits_{0 < \sigma_1 < \kappa <  \sigma_2 < 2\pi}\!\!\!\!\!\!\!\!\!\!\! d\sigma_1\, d\sigma_2\;
    \cos\of{n(\sigma_1-\sigma_2)}    
  \,,
  \nonumber\\
\end{eqnarray}
so that the divergence of the $A^3$ term is proportional, at each point $\kappa$, to
\begin{eqnarray}
  \alpha^\prime \,
  A_\mu A_\nu A_\lambda
  \left(  
    \eta^{\mu\nu}X^{\prime \lambda}\of{\kappa}
    -
    2
    \eta^{\mu\lambda}X^{\prime\nu}\of{\kappa}
    +
    \eta^{\nu\lambda}X^{\prime \mu}\of{\kappa}
  \right)
  &=&
  \alpha^\prime
  \commutator{A^\mu}{\commutator{A_\mu}{A_\lambda}}
  X^{\prime \lambda}\of{\kappa}
  \,.
  \nonumber\\
\end{eqnarray}
This vanishes precisely if
\begin{eqnarray}
  \commutator{A^\mu}{\commutator{A_\mu}{A_\lambda}} &=& 0
  \,.
\end{eqnarray}
Assuming that the restriction to constant gauge connections does not affect the generality of 
this result we hence find that at first order the condition for the well-definedness of the deformation operator
\refer{deformation expansion} is equivalent to 
\begin{eqnarray}
  \mathrm{div}_A F_A = 0
  \,,
\end{eqnarray}
which are of course just the equations of motion of Yang-Mills theory. The full equations of motion for the
nonabelian 2-form background found this way are hence
\begin{eqnarray}
  &&B = -\frac{1}{T} F_A
  \nonumber\\
  &&
  \mathrm{div}_A F_A = 0 = \mathrm{div}_A B
  \,.
\end{eqnarray}

This might appear not to be surprising. However, from the point of view of several approaches to the
topic to 2-form gauge theories found in the literature
(e.g. \cite{GirelliPfeiffer:2004,Lahiri:2001}) it may look odd, because
these equations of motion are invariant under the first order gauge transformation \refer{1st gauge trafo}
but not under that at second order \refer{2nd gauge trafo}.

But the discussion in \S\fullref{2-form gauge transformations} should clarify this: Both first and
second order gauge transformations are symmetries of the \emph{flat} 
connections on the space of (closed!) loops
and hence of the closed string. (For non-flat connections on loop space we found that no consistent
formulation is possible at all.) But due to the flatness condition the coupling of the closed string
to the nonabelian background fields is trivial, as it should be. 

The open string does couple nontrivially to the nonabelian 2-form background, but since the boundary of the
disc diagram attached to the brane couples to $A$, there is no symmetry under $A \to A + \Lambda$, and
from the heuristic picture of string physics there should not.

\section{Summary and Conclusion}
\label{Summary and conclusion}
 
We have demonstrated how a nonabelian connection on loop space 
\begin{eqnarray}
  \extd + iT \oint_A (B) = 
  \extd + it \int\limits_0^{2\pi} U_A\of{2\pi,\sigma} \commutator{\hat K}{B}\of{\sigma}U_A\of{\sigma,2\pi}
\end{eqnarray}
can be read off from certain 
deformations of the worldsheet SCFT by a generalized Wilson line along the string,
and how  formal consistency conditions on gauge transformations
of this loop space connection lead to the relation
\begin{eqnarray}
  \label{constraint in conclusion}
  \frac{1}{T}F_A + B = 0
  \,,
\end{eqnarray}
between the 1-form field strength $F_A$ and the 2-form $B$.

This consistency condition was already recently derived in
\cite{GirelliPfeiffer:2004} using the theory of 2-groups. Its role
in the theory of non-abelian surface holonomy will be discussed in detail
in \cite{Schreiber:2004g}

We have calculated the equations of motion of this nonabelian open string background by canceling 
divergences in the deformed boundary state. The result was, at lowest nontrivial order,
the ordinary equations of motion of Yang-Mills theory with respect to the 1-form connection $A$,
together with the similar equation for $B$, implied by the constraint \refer{constraint in conclusion}.

As discussed in \cite{GirelliPfeiffer:2004}, this constraint prevents 
these equations of motion to follow from a simple
generalization of the Yang-Mills action to 2-forms. Therefore precisely how the above
fits into the framwork of ``higher gauge theory'' remains to be seen. 

Conversely, the considerations presented here should show that some aspects of such 2-form gauge theories
can be understood by studying the target space theory which is \emph{implied} by certain worldsheet
theories. 

What is certainly missing, however, is a good understanding of the nature of the backgrounds described
by such worldsheet theories.

\acknowledgments{
I would like to thank 
Robert Graham, 
Christiaan Hofman, 
Martin Cederwall, 
Jens Fjelstad, 
Amitabha Lahiri,
Hendryk Pfeiffer, 
Peter Woit,
and Orlando Alvarez,
Luiz Ferreira,
Joaquin S{\'a}nchez Guill{\'e}n
and John Baez
for helpful conversation, and Charlie Stromeyer for
drawing my attention to nonabelian 2-forms in the first place.

This work was supported by SFB/TR 12
}

\newpage

\appendix

\section{Boundary state formalism}
\label{Boundary state formalism}

As a background for \S\fullref{Boundary state deformations from unitary loop space deformations}
this section summarizes  basic aspects of boundary conformal field theory 
(as discussed for instance in \cite{RecknagelSchomerus:1999,Schomerus:2002,Baumgartl:2003}).

\subsection{BCFTs}

Given a conformal field theory on the complex plane (with coordinates
$z,\bar z$) we get an associated ('descendant') 
\emph{boundary conformal field theory} (BCFT) on the upper half plane (UHP),
$\Im\of{z} > 0$, by demanding suitable boundary condition on the real line. 
The only class of cases well understood so far is that where the chiral fields
$W\of{z}, \bar W\of{\bar z}$ can be analytically continued to the real line
$\Im\of{z} = 0$ and a local automorphism of the chiral algebra exists,
the \emph{gluing map} $\Omega$, such that on the boundary the left- and right-moving
fields are related by
\begin{eqnarray}
  \label{gluing condition}
  W\of{z} &=& \Omega \bar W\of{\bar z}\,,\hspace{1cm}\mbox{at $z = \bar z$}
  \,.
\end{eqnarray}
In particular $\Omega$ always acts trivially on the energy momentum current
\begin{eqnarray}
  \Omega \bar T\of{\bar z} &=& \bar T\of{\bar z}
\end{eqnarray}
so that
\begin{eqnarray}
  T\of{z} &=& \bar T\of{\bar z}\,,\hspace{1cm}\mbox{at $z = \bar z$}
  \,,
\end{eqnarray}
which ensures that no energy-momentum flows off the boundary.

This condition allows to introduce for every chiral $W, \bar W$ the single chiral field
\begin{eqnarray}
  \mathrm{W}\of{z}
  &=&
  \left\lbrace
    \begin{array}{ll}
       W\of{z} & \mbox{for $\Im\of{z} \geq 0$}\\
       \Omega\of{\bar W}\of{\bar z} & \mbox{for $\Im\of{z} < 0$}
    \end{array}
  \right.
\end{eqnarray}
defined in the entire plane. (This is known as the 'doubling trick'.)

\subsection{Boundary states}
\label{boundary states}

Since it is relatively awkward to work with explicit constraints it is
desirable to find a framework where the boundary condition on fields at the real
line can be replaced by an operator insertion in a bulk theory without boundary.

Imagine an open string propagating with both ends attached to some D-brane. 
The worldsheet is topologically the disk (with appropriate operator insertions at the
boundary). This disk can equivalently be regarded as the half sphere glued to the
brane. But from this point of view it represents the worldsheet of a closed string
with a certain source at the brane. Therefore the open string disk correlator
on the brane is physically the same as a closed string emission from the brane with
a certain source term corresponding to the open string boundary condition.
The source term at the boundary of the half sphere can be represented 
by an operator insertion in the full sphere. The state corresponding to this vertex
insertion is the \emph{boundary state}.

In formal terms this heuristic picture translates to the following procedure:

First map the open string worldsheet to the sphere, in the above sense.
By stereographic projection, the sphere is mapped to the plane and the upper half sphere
which represents the open string worldsheet disk gets mapped to the complement of the
unit disk in the plane. Denote the complex coordinates on this complement by $\zeta, \bar \zeta$ 
and let the open string worldsheet time $\tau = -\infty$ be mappped to $\zeta=1$
and $\tau = +\infty$ mapped to $\zeta=-1$ (so that the open string propagates 'from right to left'
in these worldsheet coordinates). With $z,\bar z$ the coordinates on the UHP this corresponds to
$z = 0 \mapsto \zeta = 1$ and $z = \infty \mapsto \zeta = -1$.
The rest of the boundary of the string must get mapped to the unit circle, which is where the
string is glued to the brane. An invertible holomorphic map from the UHP to the complement of the unit disk
with these features\footnote{
  \begin{eqnarray}
    |\zeta|^2 &=& \frac{1 + |z|^2 + 2\Im\of{z}}{1 + |z|^2 - 2 \Im\of{z}} \geq 1\;\;\;
    \mbox{for $\Im\of{z} \geq 0$}
  \end{eqnarray}
} is
\begin{eqnarray}
  \zeta\of{z}
  &\defas&
  \frac{1- i z}{1 + iz}
  \,.
\end{eqnarray}

For a given boundary condition $\alpha$ the boundary state $\ket{\alpha}$ is now
defined as the state corresponding to the operator which, when inserted in the
sphere, makes the correlator of some open string field $\Phi$ 
on the sphere equal to that on the UHP with boundary condition $\alpha$:
\begin{eqnarray}
  \langle
    \Phi^{(\mathrm{H})}\of{z,\bar z}
  \rangle_\alpha
  &=&
  \left(
    \frac{\partial \zeta}{\partial z}
  \right)^{h}
  \left(
    \frac{\partial \bar \zeta}{\partial \bar z}
  \right)^{\bar h}
  \bra{0}
   \Phi^{(\mathrm{P})}\of{\zeta,\bar \zeta}
  \ket{\alpha}
  \,.
\end{eqnarray}
 
Noting that on the boundary we have
\begin{eqnarray}
  \frac{\partial \zeta}{\partial z}
  &=&
  -i \zeta
  \,,
  \hspace{.8cm}
  \mbox{at $z = \bar z \Leftrightarrow \zeta = 1/\bar \zeta$}
\end{eqnarray}
the gluing condition \refer{gluing condition} becomes
in the new coordinates
\begin{eqnarray}
  \label{gluing condition in zeta coords}
  &&\left(\frac{\partial \zeta}{\partial z}\right)^{h}
  W\of{\zeta}
  =
  \left(\frac{\partial \bar \zeta}{\partial \bar z}\right)^{h}
  \Omega \bar W\of{\bar \zeta}  
  \nonumber\\
  &\Leftrightarrow&
  W\of{\zeta}
  =
  (-1)^h \bar \zeta^{2h}\,\Omega\bar W\of{\bar \zeta}
  \,,
  \hspace{.8cm}\mbox{at $\zeta = 1/\bar \zeta$}\,.
\end{eqnarray}
In the theory living on the plane this condition translates into a constraint on the boundary state
$\ket{\alpha}$:
\begin{eqnarray}
  0 &\shallbe&
  \bra{0}
    \cdots
  \sum\limits_{n=-\infty}^\infty
  \left(
  W_n \zeta^{-n-h}
  -
  (-1)^h
  \zeta^{-2h}
  \Omega\bar W_n \zeta^{n+h}
  \right)
  \ket{\alpha}
  \nonumber\\
  &=&
  \bra{0}
    \cdots
  \sum\limits_{n=-\infty}^\infty
  \left(
  W_n \zeta^{-n-h}
  -
  (-1)^h
  \Omega\bar W_n \zeta^{n-h}
  \right)
  \ket{\alpha}
  \,,
  \hspace{.8cm}
  \forall\, \zeta = 1/\bar \zeta
  \,,
\end{eqnarray}
i.e.
\begin{eqnarray}
  \label{constraint on boundary state}
  \left(W_n - (-1)^h\Omega\bar W_{-n}\right)\ket{\alpha}
  &=& 0
  \,,
  \hspace{.8cm}
  \forall\, n\in \N
  \,.
\end{eqnarray}

Since $\Omega\bar T = \bar T$ holds for all BCFTs this implies in particular that
one always has
\begin{eqnarray}
  \label{Virasoro constraint on boundary state}
  \left(L_n - \bar L_{-n}\right)\ket{\alpha}  &=& 0\hspace{.8cm}\forall\, n
  \,,
\end{eqnarray}
which says 
that $\ket{\alpha}$ is invariant with respect to reparametrizations of the
spatial worldsheet variable $\sigma$ parameterizing the boundary
(\cf for instance section 3 of \cite{Schreiber:2004}).

\section{Computations}
\label{computations}

This section complies some calculatins which have been omitted from the main text.

\subsection{Global gauge transformations on loop space}

For completeness the following gives a derivation of \refer{1st gauge trafor from loop space}, which
is essentially nothing but the ordinary proof of gauge invariance of the Wilson line:

Consider any path-ordered Integral $I_R$ over some arbitrary object $R$ from, say 0 to 1:
\begin{eqnarray}
  I_R &\defas&
  \lim\limits_{N = 1/\epsilon \to \infty}  
  \left[
    \left(
      1 + \epsilon R\of{0}
    \right)
    \left(
      1 + \epsilon R\of{\epsilon}
    \right)
    \left(
      1 + \epsilon R\of{2\epsilon}
    \right)
    \cdots
    \left(
      1 + \epsilon R\of{1}
    \right)
  \right]
  \,.
  \nonumber\\
\end{eqnarray}
Let $U = U\of{\sigma}$ be a unitary function on the loop which does not depend on the 
embedding field $X$. By acting with it on the endpoints of the
above path ordered integral we get
\begin{eqnarray}
  U\of{0}I_RU^\dagger\of{1} 
  &\defas&
  \lim\limits_{N = 1/\epsilon \to \infty}  
  U\of{0}
  \left[
    \left(
      1 + \epsilon R\of{0}
    \right)
    U^\dagger\of{\epsilon}U\of{\epsilon}
    \left(
      1 + \epsilon R\of{\epsilon}
    \right)
    U^\dagger\of{2\epsilon}U\of{2\epsilon}
    \cdots
    \left(
      1 + \epsilon R\of{1}
    \right)
  \right]
  U^\dagger\of{1}
  \nonumber\\
  &=&
  \lim\limits_{N = 1/\epsilon \to \infty}  
  \Big[
    \left(
      U\of{0}U^\dagger\of{\epsilon} + \epsilon U\of{0} R\of{0}U^\dagger\of{\epsilon}
    \right)
    \cdots
    \nonumber\\
    && 
  \hspace{50pt}\cdots
    \left(
      U\of{(N-1)\epsilon}U^\dagger\of{1} + \epsilon U\of{(N-1)\epsilon}R\of{1}U^\dagger\of{1}
    \right)
  \Big]
  \nonumber\\
  &=&
  \lim\limits_{N = 1/\epsilon \to \infty}  
  \left[
    \left(
      1 + \epsilon UU^{\dagger \prime} + \epsilon U RU^\dagger
    \right)\of{0}
    \cdots
    \left(
      1 + \epsilon UU^{\dagger \prime} + \epsilon U RU^\dagger
    \right)\of{1}
  \right]
  \,.
\end{eqnarray}

If the term in $R$ proportional to $X^{\prime\mu}$ is identified with $A_\mu$ this gives the
gauge transformation $A \mapsto U(dU^\dagger) + UAU^\dagger$ for $A$ and $B \mapsto UBU^\dagger$
for the remaining components of $R$.

The same applies in the other $\sigma$-direction:
\begin{eqnarray}
  U\of{1}J_R U^\dagger\of{0} 
  &\defas&
  \lim\limits_{N = 1/\epsilon \to \infty}  
  U\of{1}
  \Big[
    \left(
      1 - \epsilon R\of{1}
    \right)
    U^\dagger\of{(N-1)\epsilon}U\of{(N-1)\epsilon}
    \nonumber\\
    &&
    \hspace{50pt}   
    \cdots
    \left(
      1 + \epsilon R\of{\epsilon}
    \right)
    U^\dagger\of{(N-2)\epsilon}U\of{(N-2)\epsilon}
    \left(
      1 - \epsilon R\of{0}
    \right)
  \Big]
  U^\dagger\of{0}
  \nonumber\\
  &=&
  \lim\limits_{N = 1/\epsilon \to \infty}  
  \Big[
    \left(
      U\of{1}U^\dagger\of{(N-1)\epsilon} - \epsilon U\of{1} R\of{1}U^\dagger\of{(N-1)\epsilon}
    \right)
    \cdots
    \nonumber\\
    && 
  \hspace{50pt}\cdots
    \left(
      U\of{\epsilon}U^\dagger\of{0} - \epsilon U\of{\epsilon}R\of{0}U^\dagger\of{0}
    \right)
  \Big]
  \nonumber\\
  &=&
  \lim\limits_{N = 1/\epsilon \to \infty}  
  \left[
    \left(
      1 - \epsilon UU^{\dagger \prime} - \epsilon U RU^\dagger
    \right)\of{1}
    \cdots
    \left(
      1 - \epsilon UU^{\dagger \prime} - \epsilon U RU^\dagger
    \right)\of{0}
  \right]
  \,.
  \nonumber\\
\end{eqnarray}

In particular, it follows that
\begin{eqnarray}
  U\of{0}
  \left(
    \extd
    +
    \oint_A (B)
  \right)
  U^\dagger\of{0}
  &=& 
  \extd
  +
  \oint_{UAU^\dagger + U(U^\dagger)^\prime}
  (UBU^\dagger)
  \,.
\end{eqnarray}

This directly gives equation \refer{1st gauge trafor from loop space} used in the main text.

\subsection{The exterior derivative of path-ordered loop space forms}
 \label{The exterior derivative of path-ordered loop space forms}

The action of the loop space exterior derivative in 
\refer{extd on path ordered forms}  is derived as follows:
\begin{eqnarray}
  &&\commutator{\extd}{\oint (\omega_1, \cdots, \omega_n)}
  \nonumber\\
  &=&
  \sum\limits_k (-1)^{\left(1+\sum\limits_{i < k}p_i\right)}
  \oint (\omega_1, \cdots, d\omega_k, \cdots, \omega_n)
  \nonumber\\
  &&
  +
  \sum\limits_k (-1)^{\left(\sum\limits_{i < k}p_i\right)}
  \int\limits_{0 < \sigma_{i-1} < \sigma_i < \sigma_{i+1} < \pi}
  \!\!\!\!\!\!\!
  d^n\sigma\;
  \commutator{\hat K}{\omega_1}\of{\sigma_1}
  \cdots
  \left(\omega_k\right)^\prime
  \cdots
  \commutator{\hat K}{\omega_n}\of{\sigma_n}
  \nonumber\\
  &=&
  \sum\limits_k (-1)^{\left(1+\sum\limits_{i < k}p_i\right)}
  \oint (\omega_1, \cdots, d\omega_k, \cdots, \omega_n)
  \nonumber\\
  &&
  +
  \sum\limits_k (-1)^{\left(1+\sum\limits_{i < k}p_i\right)}
  \nonumber\\
  &&
  \int\limits_{0 < \sigma_{i-1} < \sigma_i < \sigma_{i+1} < \pi}
  \!\!\!\!\!\!\!
  d^n\sigma\;
  \commutator{\hat K}{\omega_1}\of{\sigma_1}
  \cdots
  \left(
   \commutator{\hat K}{\omega_{k-1}}
    \omega_{k}
    -
    (-1)^{p_{k-1}}
    \omega_{k-1}
    \commutator{\hat K}{\omega_{k}}
  \right)
\of{\sigma_k}
  \cdots
  \commutator{\hat K}{\omega_n}\of{\sigma_{n-1}}  
  \nonumber\\
  &=&
  \sum\limits_k (-1)^{\left(1+\sum\limits_{i < k}p_i\right)}
  \left(
    \oint (\omega_1, \cdots, d\omega_k, \cdots, \omega_n)
    +
    \oint (\omega_1,\cdots, \omega_{k-1}\wedge \omega_k,\cdots,\omega_n)
  \right)
  \,.
  \nonumber
\end{eqnarray}

\newpage

\bibliography{std}

\end{document}